\documentclass[english, openright, 12pt]{book}%
\usepackage[utf8]{inputenc}
\usepackage[TS1,T1]{fontenc}
\usepackage{textcomp}
\usepackage{mathptmx}

\usepackage[bookmarksnumbered=true]{hyperref}

\usepackage{xcolor}
\usepackage{tikz}
\usepackage{graphicx}
\usepackage{bm}
\usepackage{amssymb}
\usepackage{verbatim}
\usepackage{amsmath,amsfonts,amssymb}

\usepackage{caption}

\newcommand{\beqa}{\begin{eqnarray}}
\newcommand{\eeqa}[1]{\label{#1}\end{eqnarray}}
\newcommand{\beq}{\begin{equation}}
\newcommand{\eeq}{\end{equation}}

\newcommand{\ba}{\begin{eqnarray}}
\newcommand{\ea}{\end{eqnarray}}

\def\R{{\rm I\!R}}

\usepackage{bm}
\def\bxi {{\bm \xi}}

\def\dint {\int\int}

\makeatletter
\renewcommand\section{\@startsection {section}{1}{\z@}%
	{-3.5ex \@plus -1ex \@minus -.2ex}%
	{2.3ex \@plus .2ex}%
	{\reset@font\bfseries}}
\renewcommand\subsection{\@startsection {subsection}{1}{\z@}%
	{-3.5ex \@plus -1ex \@minus -.2ex}%
	{2.3ex \@plus .2ex}%
	{\reset@font\it \bfseries}}
\renewcommand\subsubsection{\@startsection {subsubsection}{1}{\z@}%
	{-3.5ex \@plus -1ex \@minus -.2ex}%
	{2.3ex \@plus .2ex}%
	{\reset@font\it \bfseries \underline}}
\makeatother

\usepackage[a4paper,tmargin=2.5cm,bmargin=3.0cm,rmargin=2.5cm,lmargin=2.5cm]{geometry} 
\usepackage{caption}

\usepackage{fancyhdr}
\begin{document}

\let\cleardoublepage\clearpage
\renewcommand\contentsname{\normalsize{\hspace{12 pt}Table of Contents:}}
\tableofcontents
\thispagestyle{empty}
\let\cleardoublepage\clearpage
\newpage

\pagenumbering{arabic}
\newcommand{\numchapter}{11} 
\newcommand{\firstpage}{1} 
\setcounter{page}{\firstpage} 
\setcounter{chapter}{\numchapter} 

\cfoot{}
\begin{center}
{\fontsize{20}{20}\selectfont \textbf{
\hfill Chapter \numchapter\\[5mm]
Homogenization Techniques for Periodic Structures\\[10mm] 
}}
{\fontsize{14}{14}\selectfont Sebastien Guenneau$^{(1)}$, Richard Craster$^{(2)}$, Tryfon Antonakakis$^{(2,3)}$,
Kirill Cherednichenko$^{(4)}$ and Shane Cooper$^{(4)}$\\[5mm] } 
{\fontsize{10}{10}\selectfont \textit{ 
$^{(1)}$CNRS, Aix-Marseille Universit{\'e}, \'Ecole Centrale Marseille, Institut Fresnel,\\
13397 Marseille Cedex 20, France, {\color{blue}{\underline{sebastien.guenneau@fresnel.fr}}}\\
$^{(2)}$ Department of Mathematics, Imperial College London, United Kingdom,
{\color{blue}{\underline{r.craster@imperial.ac.uk}}},\\
$^{(3)}$ CERN, Geneva, Switzerland,{\color{blue}{\underline{tryfon.antonakakis09@imperial.ac.uk}}},\\
$^{(4)}$ Cardiff School of Mathematics, Cardiff University, United Kingdom, \\
{\color{blue}{\underline{cherednichenko@cardiff.ac.uk}}},
{\color{blue}{\underline{coopersa@cf.ac.uk}}}. 
}}
\end{center}

\thispagestyle{empty}

\pagestyle{fancyplain} 
\renewcommand{\headrulewidth}{0.0pt} 
\lhead[\itshape{\small{\thechapter .\thepage}}]{\itshape{\small{S. Guenneau et al.: Homogenization Techniques for Periodic Structures}}}
\rhead[\itshape{\small{Gratings: Theory and Numeric Applications, 2012}}]{\itshape{\small{\thechapter .\thepage}}}
\cfoot{}

\section{Introduction}
In this chapter we describe a selection of mathematical techniques and results that suggest interesting links between the theory of gratings and the theory of homogenization, including a brief introduction to the latter.  By no means do we purport to imply that homogenization theory is an exclusive method for studying gratings, neither do we hope to be exhaustive in our choice of topics within the subject of homogenization. Our preferences here are motivated most of all by our own latest research, and by our outlook to the future interactions between these two subjects. We have also attempted, in what follows, to contrast the ``classical'' homogenization (Section \ref{clashom1}), which is well suited for the description of composites as we have known them since their advent until about a decade ago, and the ``non-standard'' approaches,
high-frequency homogenization (Section \ref{hfh}) and high-contrast homogenization (Section \ref{kc}),
which have been developing in close relation to the study of photonic crystals and metamaterials, which exhibit properties unseen in conventional composite media, such as negative refraction allowing for super-lensing through a flat heterogeneous lens,
and cloaking, which considerably reduces the scattering by finite size objects (invisibility) in certain frequency range. These novel
electromagnetic paradigms have renewed the interest of physicists and applied mathematicians alike in the theory of gratings
\cite{bookjaune}.

\subsection{Historical survey on homogenization theory}
The development of theoretical physics and continuum mechanics in the second half of the 19th and first half 
of the 20th century has 
motivated the question of justifying the macrosopic view of physical phenomena (at the scales visible to the 
human eye) by ``upscaling'' the implied microscopic rules for particle interaction at the atomic level through 
the phenomena at the intermediate, ``mesoscopic'', level (from tenths to hundreds of microns). This ambition 
has led to an extensive worldwide programme of research, which is still far from being complete as of now. 
Trying to give a very crude, but more or less universally applicable, approximation of the aim of this extensive activity, one could say that it has to do with developing approaches to averaging out in some way material properties at one level with the aim of getting a less detailed, but almost equally precise, description of the material response. Almost every word in the last sentence needs to be clarified already, and this is essentially the point where one could start giving an overview of the activities that took place during the years to follow the great physics advances of a century ago. Here we focus on the research that has been generally referred to as the theory of homogenization, starting from the early 1970s. Of course, even at that point it was not, strictly speaking, the beginning of the 
subject, but we will use this period as a kind of reference point in this survey.
   
   The question that a mathematician may pose in relation to the perceived concept of ``averaging out'' the detailed features of a heterogeneous structure in order to get a more homogeneous description of its behaviour is the following: suppose that we have the simplest possible linear elliptic partial differential equation (PDE) with periodic coefficients of period 
   $\eta>0.$ What is the asymptotic behaviour of the solutions to this PDE as $\eta\to0$? Can a boundary-value problem be written that is satisfied by the leading term in the asymptotics, no matter what 
   the data unrelated to material properties are? Several research groups became engaged in addressing this question about four decades ago, most notably those led by N. S. Bakhvalov, E. De Giorgi, J.-L. Lions, V. A. Marchenko, see \cite{Bakhvalov}, \cite{DeGiorgi}, \cite{Bensoussan78a}, \cite{MarchenkoKhruslov} for some of the key contributions of 
   that period. The work of these groups has immediately led to a number of different perspectives on the apparently basic question asked above, which in part was due to the different contexts that these research groups had 
   had exposure to prior to dealing with the issue of averaging. Among these are the method of multiscale 
   asymptotic expansions (also discussed later in this chapter), the ideas of compensated compactness 
   (where the contribution by L. Tartar and F. Murat \cite{Tartar}, \cite{Murat} has to be mentioned specifically), 
   the variational method (also known as the ``$\Gamma$-convergence"). These approaches were subsequently 
   applied to various contexts, both across a range of mathematical setups (minimisation problems, hyperbolic equations, problems with singular boundaries)  and across a number of physical contexts (elasticity, electromagnetism, heat conduction). Some new approaches to homogenization appeared later on, too, such as 
   the method of two-scale convergence by G. Nguetseng \cite{Nguetseng}  and the periodic 
   unfolding technique by D. Cioranescu, A. Damlamian and G. Griso \cite{CDG}. 
   Established textbooks that summarise these developments in different time periods, include, in 
   addition to the already cited book \cite{Bensoussan78a}, the monographs \cite{BP}, \cite{jikov94a}, \cite{SanchezPalencia}, 
   and more recently \cite{ChPSh}. The area that is perhaps worth a separate mention is that of stochastic homogenization, where some pioneering contributions were made by S. M. Kozlov \cite{Kozlov}, G. C. Papanicolaou and S. R. S. Varadhan \cite{PV}, and which has in recent years been approached 
   with renewed interest.
    
    A specific area of interest within the subject of homogenization that has been rapidly developing during the last decade or so is the study of the behaviour of "non-classical" periodic structures, which we understand here as 
    those for which compactness of bounded-energy solution sequences fails to hold as $\eta\to0.$ The related mathematical research has been strongly linked to, and indeed influenced by, the parallel development of the 
    area of metamaterials and their application in physics, in particular for electromagnetic phenomena. 
    Metamaterials can be roughly defined as those whose properties at the macroscale are affected by 
    higher-order behaviour as $\eta\to0.$ For example, in classical homogenization for elliptic second-order PDE one requires the leading (``homogenised solution'') and the first-order (``corrector'') terms in the $\eta$-power-series expansion of the solution in order to determine the  macroscopic properties, which results in a limit of the same type as the original problem, where the solution flux  (``stress'' in elasticity, ``induction'' in electromagnetics, ``current'' in electric conductivity, 
    ``heat flux'' in heat conduction) depends on the solution gradient only (``strain'' in elasticity, 
    "field" in electromagnetics, ``voltage'' in electric conductivity, ``temperature gradient'' in heat condiction). 
    If, however, one decides for some reason, or is forced by the specific problem setup, to include higher-order terms as well, they are likely to have to deal with an asymptotic limit of a different type for small $\eta,$ which may, say, include second gradients of the solution in its constitutive law. One possible reason for the need to include such unusual effects is the non-uniform  (in $\eta$) ellipticity  of the original problems or, using the language of materials science, the high-contrast in the material properties of the given periodic structure.  Perhaps the earliest mathematical example of such degeneration is the so-called "double-porosity model", which was first considered by G. Allaire \cite{Allaire} and T. Arbogast, J. Douglas, U. Hornung \cite{Arbogastetal} in the early 1990s. A detailed analysis of the properties of double-porosity models, including their striking spectral behaviour did not appear until the work  \cite{Zhikov2000} by V. V. Zhikov. We discuss the double-porosity model and its properties in more detail in Section \ref{kc}.
     
     Before moving on to the next section, it is important to mention one line of research within the homogenization area that has had a significant r\^{o}le in terms of application of mathematical analysis to materials, namely the subject   of  periodic singular structures (or ``multi-structures'', see \cite{KMM}). While this subject is clearly linked to the general analysis of differential operators on singular domains (see \cite{MNP}), there has been a series of works that develop specifically homogenization techniques for periodic structures of this kind  (also referred to as ``thin structures'' in this context), {\it e.g.} \cite{Zhikovstructures}, \cite{ZhikovPastukhova}. It turns out that overall properties of such materials are similar to those of materials with high contrast. In the same vein, it is not difficult to see that compactness of bounded-energy sequences for problems on periodic thin structures does not hold (unless the sequence in question is suitably rescaled), which leads to the need for non-classical, higher-order, techniques in their analysis.
 
\subsection{Multiple scale method: Homogenization of microstructured fibers}\label{clashom1}

\begin{figure}[h]\centerline{\scalebox{0.7}{\includegraphics{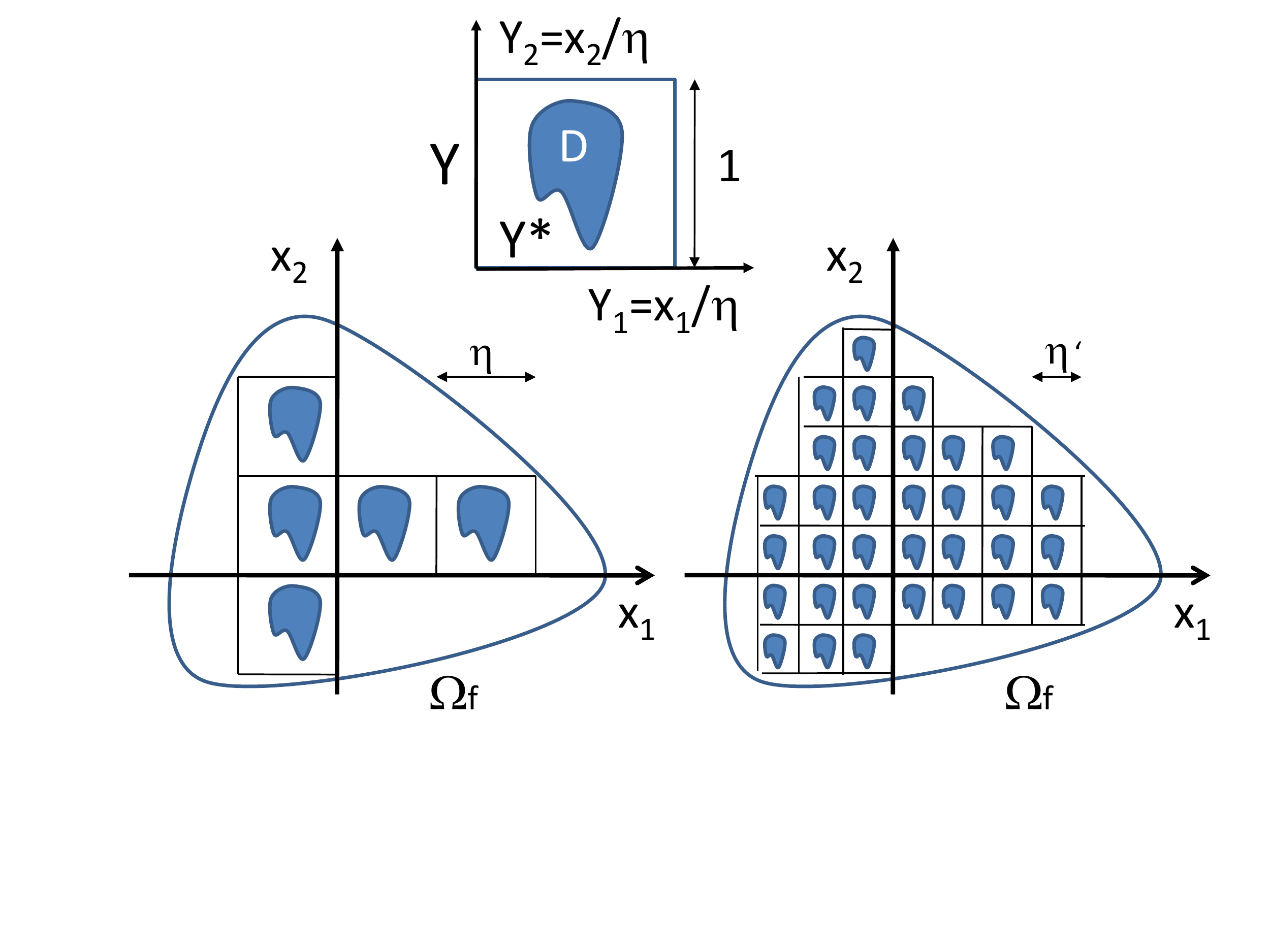}}}
\vspace{-2cm}\caption{A diagram of the homogenization process:
when the parameter $\eta$ gets smaller ($\eta<\eta'$), the number of cells inside the fixed domain $\Omega_f$ becomes
larger. When $\eta\ll 1$, $\Omega_f$ is filled with a large number of small cells, and can thus be considered as an
effective (or homogenized) medium. Such a medium is usually described by anisotropic parameters depending upon
the resolution of auxiliary (``unit cell'') problems set on the rescaled microcopic cell $Y$ which typically contains one inclusion $D$.}
  \label{fig1}
  \end{figure}

Let us consider a doubly periodic grating of pitch $\eta$ and finite extent such as shown in Fig.\ref{fig1}. An interesting problem to look at is that of transverse electric (TE) modes--- when the magnetic field has the form $(0,0,H)$---
propagating within a micro-structured fiber with infinite conducting walls. Such an eigenvalue problem is known to have
a discrete spectrum: we look for eigenfrequencies $\omega$ and associated eigenfields $H$ such that:
$$
({\cal P}_\eta): \left\{
\begin{array}{ll}
\displaystyle{-\sum_{i,j=1}^2\frac{\partial}{\partial x_i}\left( \varepsilon_{ij}^{-1}(\frac{{\bf x}}{\eta})
\frac{\partial H({\bf x})}{\partial x_j}\right)} = \omega^2 \mu_0\varepsilon_0 H({\bf x})  \; & \hbox{in $\Omega_f$} \; , \\
\displaystyle{\varepsilon_{ij}^{-1}(\frac{{\bf x}}{\eta})
\frac{\partial H({\bf x})}{\partial x_i}n_j}=0 \; & \hbox{on $\partial\Omega_f$} \; ,
\end{array}
\right.
$$
where we use the convention ${\bf x}=(x_1,x_2)$, $\partial\Omega_f$ denotes the boundary $\Omega_f$,
and ${\bf n}=(n_1, n_2)$ is the normal to the boundary.
Here, $\varepsilon_0\mu_0=c^{-2}$ where $c$ is the speed of light in vacuum
and we assume that matrix coefficients of relative permittivity $\varepsilon_{ij}({\bf y})$,
with $i,j=1,2$, are real, symmetric (with
the convention ${\bf y}=(y_1,y_2)$), of period $1$ (in $y_1$ et $y_2$) and satisfy:
\begin{equation}
M{\mid{\bm \xi}\mid}^2\geq\varepsilon_{ij}({\bf y})\xi_i\xi_j\geq m{\mid{\bm \xi}\mid}^2 \; , \; \forall {\bm \xi}\in\R^2 \; , \; \forall {\bf y}\in Y={[0,1]}^2 \; ,
\label{ineqhom}
\end{equation}
where ${\mid{\bm \xi}\mid}^2=(\xi_1^2+\xi_2^2)$, for given strictly positive constants $M$ and $m$.
This condition is met for all conventional dielectric media\footnote{When the periodic medium is assumed to be isotropic,
$\varepsilon_{ij}(y)=\varepsilon({\bf y})\delta_{ij}$, with the Kronecker symbol $\delta_{ij}=1$ if $i=j$
and $0$ otherwise.
For instance, (\ref{ineqhom}) has typically the bounds $M=13$ and $m=1$ in optics.
One class of problems where this condition (\ref{ineqhom}) is violated
(the bound below, to be more precise)
is considered in Section \ref{kc} on high-contrast homogenization.}.

\noindent We can recast $({\cal P}_\eta)$ as follows:
$$-\frac{\partial}{\partial x_i}\sigma^i(H({\bf x}))=\frac{\omega^2}{c^2}H({\bf x})$$
with
$$\sigma^i(H({\bf x}))=\varepsilon_{ij}^{-1}\left(\frac{{\bf x}}{\eta}\right)
\frac{\partial H({\bf x})}{\partial x_j} \; .$$

\noindent The multiscale method relies upon the following ansatz:
\begin{equation}
H=H_0({\bf x})+\eta H_1({\bf x},{\bf y})+\eta^2 H_2({\bf x},{\bf y}) + ...
\label{sebeq01}
\end{equation}
where $H_i({\bf x},{\bf y})$, $i=1,2,...$ is a periodic function of period $Y$ in ${\bf y}$.

\noindent In order to proceed with the asymptotic algorithm, one needs to rescale
the differential operator as follows
\begin{equation}
\frac{\partial H}{\partial x_i}=\left( \frac{\partial H_0}{\partial z_i}+\frac{\partial H_1}{\partial y_i}\right)
+\eta \left( \frac{\partial H_1}{\partial z_i}+\frac{\partial H_2}{\partial y_i}\right)+...
\label{sebeq02}
\end{equation}
where $\partial/\partial z_i$ stands for the partial derivative with respect to
the $i$th component of the macroscopic variable ${\bf x}$.

\noindent It is useful to set

$$\sigma^i(H)=\sigma^i_0+\eta\sigma^i_1+\eta^2\sigma^i_2+...$$
what makes (\ref{sebeq02}) more compact.

\noindent Collecting coefficients sitting in front of the same powers of $\eta$, we obtain:

$$\sigma^i_0(H)=\varepsilon_{ij}^{-1}({\bf y})\left( \frac{\partial H_0}{\partial z_i}+\frac{\partial H_1}{\partial y_i}\right)$$

$$\sigma^i_1(H)=\varepsilon_{ij}^{-1}({\bf y})\left( \frac{\partial H_1}{\partial z_i}+\frac{\partial H_2}{\partial y_i}\right)$$
and so forth, all terms being periodic in ${\bf y}$ of period $1$.

\noindent Upon inspection of problem $({\cal P}_\eta)$, we gather that
$$-\left(\frac{1}{\eta}\frac{\partial}{\partial y_i}+\frac{\partial}{\partial z_i}\right)\left(\sigma^i_0+\eta\sigma^i_1+...\right)=\frac{\omega^2}{c^2}H({\bf x})+...$$
so that at order $\eta^{-1}$
$$({\cal A}):-\frac{\partial}{\partial y_i}\sigma^i_0=0 \; ,$$
and at order $\eta^0$
$$({\cal H}):-\frac{\partial}{\partial z_i}\sigma^i_0-\frac{\partial}{\partial y_i}\sigma^i_1=\frac{\omega^2}{c^2} H_0 \; .$$
(the equations corresponding to higher orders in $\eta$ will not be used here).

\noindent Let us show that $({\cal H})$ provides us with an equation (known as the homogenized equation) associated with the macroscopic
behaviour of the microstructured fiber. Its coefficients will be obtained thanks to $({\cal A})$ which is an auxiliary problem related to
the microscopic scale. We will therefore be able to compute $H_0$ and $H_1$ thus, in particular, the first terms
of $H$ and $\sigma^i$.

\noindent In order to do so, let us introduce the mean on $Y$, which we denote $<.>$, which is an operator acting on
the function $g$ of the variable ${\bf y}$:
$$<g>=\frac{1}{\mid Y\mid}\int\int_Y g(y_1,y_2) dy_1dy_2 \; ,$$
where $\mid Y \mid$ is the area of the cell $Y$.

\noindent Applying the mean to both sides of $({\cal H})$, we obtain:
$$<({\cal H})>:-\frac{\partial}{\partial z_i} <\sigma^i_0>
-<\frac{\partial}{\partial y_i}\sigma^i_1>=\frac{\omega^2}{c^2} H_0 <1> \; ,$$
where we have used the fact that $<.>$ commutes with $\partial/\partial z_i$.

\noindent Moreover, invoking the divergence theorem, we observe that
$$<\frac{\partial}{\partial y_i}\sigma^i_1>=\frac{1}{\mid Y\mid}\int\int_Y \frac{\partial}{\partial y_i}\sigma^i_1({\bf y}) d{\bf y}
=\frac{1}{\mid Y\mid}\int_{\partial Y} \sigma^i_1({\bf y})n_i ds \; ,$$
where ${\bf n}=(n_1,n_2)$ is the unit outside normal to $\partial Y$ of $Y$.
This normal takes opposite values on opposite sides of $Y$, hence the integral over $\partial Y$ vanishes.

\noindent Altogether, we obtain:
$$<({\cal H})>:-\frac{\partial}{\partial z_i} <\sigma^i_0> =\frac{\omega^2}{c^2} H_0 \; ,$$
which only involves the macroscopic variable $x$ and partial derivatives
$\partial /\partial z_i$ with respect to the macroscopic variable.
We now want to find a relation between
$<\sigma_0>$ and the gradient in ${\bf x}$ of $H_0$. Indeed, we have seen that
$$\sigma^i_0(H)
=\varepsilon_{ij}^{-1}({\bf y})\left( \frac{\partial H_0}{\partial z_j}+\frac{\partial H_1}{\partial y_j}\right) \; ,$$

\noindent which from $({\cal A})$ leads to
$$({\cal A}1):-\frac{\partial}{\partial y_i}\left(\varepsilon_{ij}^{-1}({\bf y})\frac{\partial H_1}{\partial y_j}\right)
=\left( \frac{\partial H_0}{\partial z_j}\right) \left(\frac{\partial}{\partial y_i}\varepsilon_{ij}^{-1}({\bf y})\right) \; .$$

\noindent We can look at $({\cal A}1)$ as an equation for the unknown $H_1({\bf x},{\bf y})$, periodic of period $Y$ in ${\bf y}$
and parametrized by ${\bf x}$. Such an equation is solved up to an additive constant.
In addition to that, the parameter ${\bf x}$ is only involved via the factor
$\partial H_0/\partial z_j$. Hence, by linearity, we can write the solution $H_1({\bf x},{\bf y})$ as follows:
$$H_1({\bf x},{\bf y})= \frac{\partial H_0({\bf x})}{\partial z_j} w^j({\bf y}) \; ,$$
where the two functions $w^j({\bf y})$, $j=1,2$ are solutions to $({\cal A}1)$ corresponding to
$\partial H_0/\partial z_j({\bf x})$, $j=1,2$ equal to unity with the other ones being zero, that is solutions to:
$$({\cal A}2):-\frac{\partial}{\partial y_i}\left(\varepsilon_{ij}^{-1}({\bf y})\frac{\partial w^k}{\partial y_j}\right)
=\delta_{jk} \left(\frac{\partial}{\partial y_i}\varepsilon_{ij}^{-1}({\bf y})\right) \; ,$$
with $w^k({\bf y})$, $k=1,2$ periodic functions in ${\bf y}$ of period $Y$
\footnote{We note that $({\cal A}2)$ are two equations which merely depend upon $\varepsilon_{ij}^{-1}({\bf y})$, that is
on the microscopic properties of the periodic medium. The two functions $w^k$ (defined up to an
additive constant) can be computed once for all, independently of $\Omega_f$.}.

\noindent Since the functions $w^k({\bf y})$ are known, we note that
$$\sigma_i^0({\bf x},{\bf y})=\varepsilon_{ij}^{-1}({\bf y})\left( \frac{\partial H_0}{\partial z_j}+\frac{\partial H_1}{\partial y_j}\right)
=\varepsilon_{ij}^{-1}({\bf y})\left( \frac{\partial H_0}{\partial z_j}
+\frac{\partial H_0}{\partial z_k}\frac{\partial w^k({\bf y})}{\partial y_j}\right) \; ,$$
which can be written as
$$\sigma^i_0({\bf x},{\bf y})=\left( \varepsilon_{ik}^{-1}({\bf y})
+\varepsilon_{ij}^{-1}({\bf y})\frac{\partial w^k({\bf y})}{\partial y_j}\right)\frac{\partial H_0({\bf x})}{\partial z_k} \; .$$

\noindent Lets us now apply the mean to both sides of this equation. We obtain:
$$<\sigma^i_0>({\bf x})=\varepsilon_{{\rm{hom}},ik}^{-1} \frac{\partial H_0({\bf x})}{\partial z_k} \; ,$$
which can be recast as the following homogenized problem:

$$
({\cal P}_0): \left\{\begin{array}{ll}
\displaystyle{-\sum_{i,k=1}^2\frac{\partial}{\partial x_i}\left( \varepsilon_{{\rm{hom}},ik}^{-1}
\frac{\partial H_0({\bf x})}{\partial x_k}\right)} = \omega^2 \mu_0\varepsilon_0 H_0({\bf x})  \; &, \hbox{in $\Omega_f$} \; , \\
\displaystyle{\varepsilon_{{\rm{hom}},ik}^{-1}(\frac{{\bf x}}{\eta})
\frac{\partial H_0({\bf x})}{\partial x_i}n_k}=0 \; &, \hbox{on $\partial\Omega_f$} \; ,
\end{array}
\right.
$$
where $\varepsilon_{{\rm{hom}},ik}^{-1}$ denote the coefficients of the homogenized matrix of permittivity
given by:
\begin{equation}
\varepsilon_{{\rm{hom}},ik}^{-1}=\frac{1}{\mid Y \mid}\int\int_Y\left( \varepsilon_{ik}^{-1}({\bf y})
+\varepsilon_{ij}^{-1}({\bf y})\frac{\partial w^k({\bf y})}{\partial y_j}\right) \, d{\bf y} \; .
\label{sebeq03}
\end{equation}

\noindent As an illustrative example for this homogenized problem, we consider a
microstructured waveguide consisting of a medium with relative
permittivity $\varepsilon=1.25$ with elliptic inclusions 
(of minor and major axes $0.3$ cm and $0.4$ cm respectively) with center to center spacing
$d=0.1cm$ with an infinite conducting boundary {\it i.e.} Neumann boundary
conditions in the TE polarization.

We use the COMSOL MULTIPHYSICS finite element package to solve the annex problem and we
find that ${[\varepsilon_{\rm hom}]}$ from (\ref{sebeq03}) writes as \cite{zolgue1}
$$
\left(
\begin{array}{cc}
1.9296204& -1.0533083 \, 10^{-16} \\
-44.417444 \, 10^{-18} & 2.1127643
\end{array}
\right) \; ,
$$
with $<\varepsilon>_{Y}= 2.2867255$. The off diagonal
terms can be neglected.

If we assume that the transverse propagating modes in the metallic waveguide
have a small propagation constant $\gamma\ll 1$, the above mathematical
model describes accurately the physics. We show in Fig. \ref{anisotropy}
a comparison between two TE modes of the microstructured waveguide
and its associated anisotropic homogenized counterpart. Both
eigenfrequencies and eigenfields match well (note that we use the
waveguide terminology wavenumber $k=\sqrt{\omega^2/c^2-\gamma^2}$).

\begin{figure}
\begin{minipage}[b]{0.5 \textwidth}
\centering
\begin{picture}(0,100)
\put(-100,0){\includegraphics[angle=0,width=0.8 \textwidth,
  draft=false]{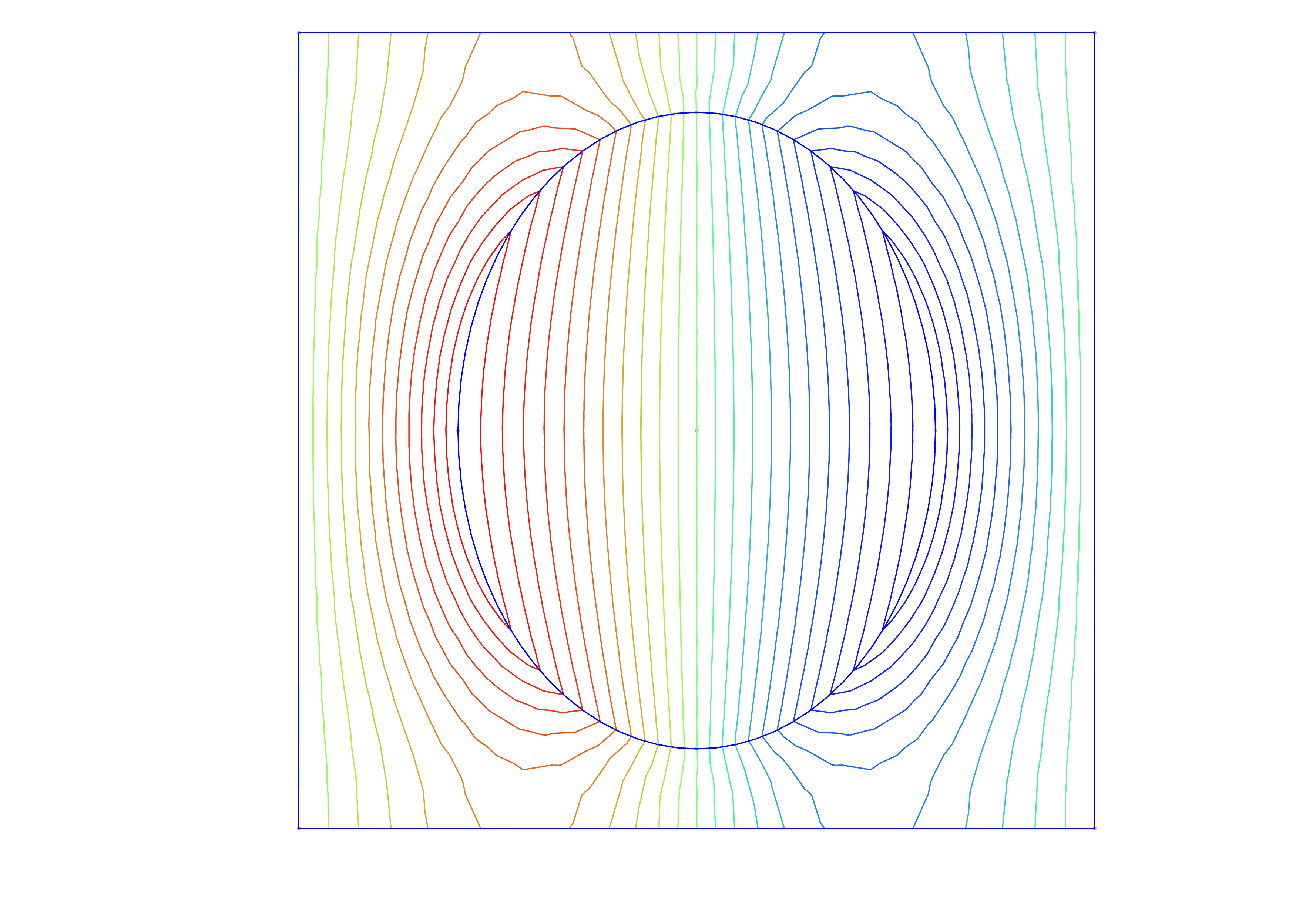}}
\end{picture}
\end{minipage}%
\begin{minipage}[b]{0.5 \textwidth}
\centering        \includegraphics[angle=0,width=0.8\textwidth,
draft=false]{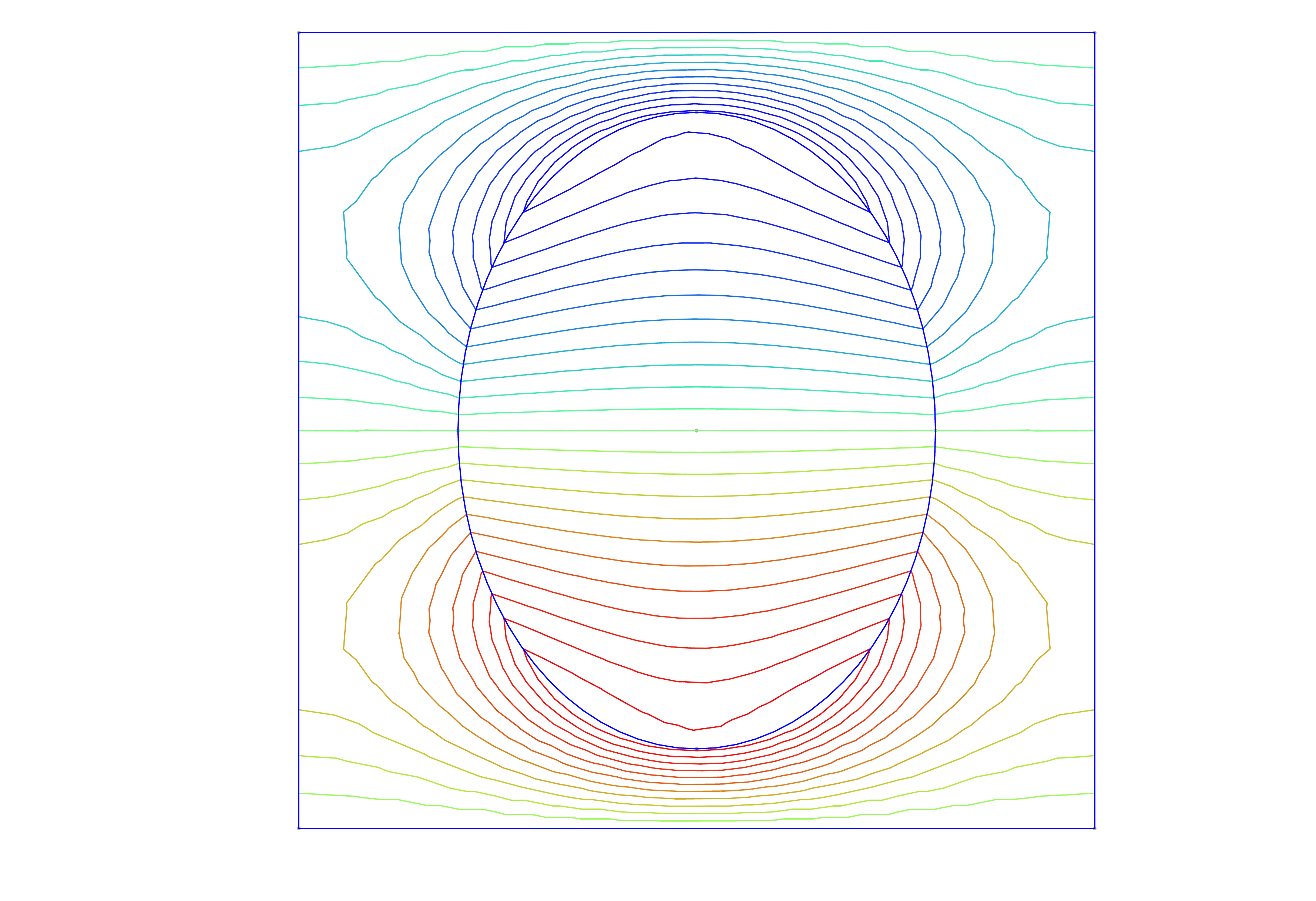}
\end{minipage}
\caption{Potentials $V_x$ (left) and $V_y$ (right):
The unit cell contains an elliptic inclusion of relative
permittivity ($\varepsilon=4.0+3i$) with minor
and major axis $a=0.3$ and $b=0.4$ in silica
$(\varepsilon=1.25)$.}
\label{potential}
\end{figure}

\begin{figure}[h!]
        \centerline{\resizebox{14cm}{!}{
    \rotatebox{-90}{\includegraphics{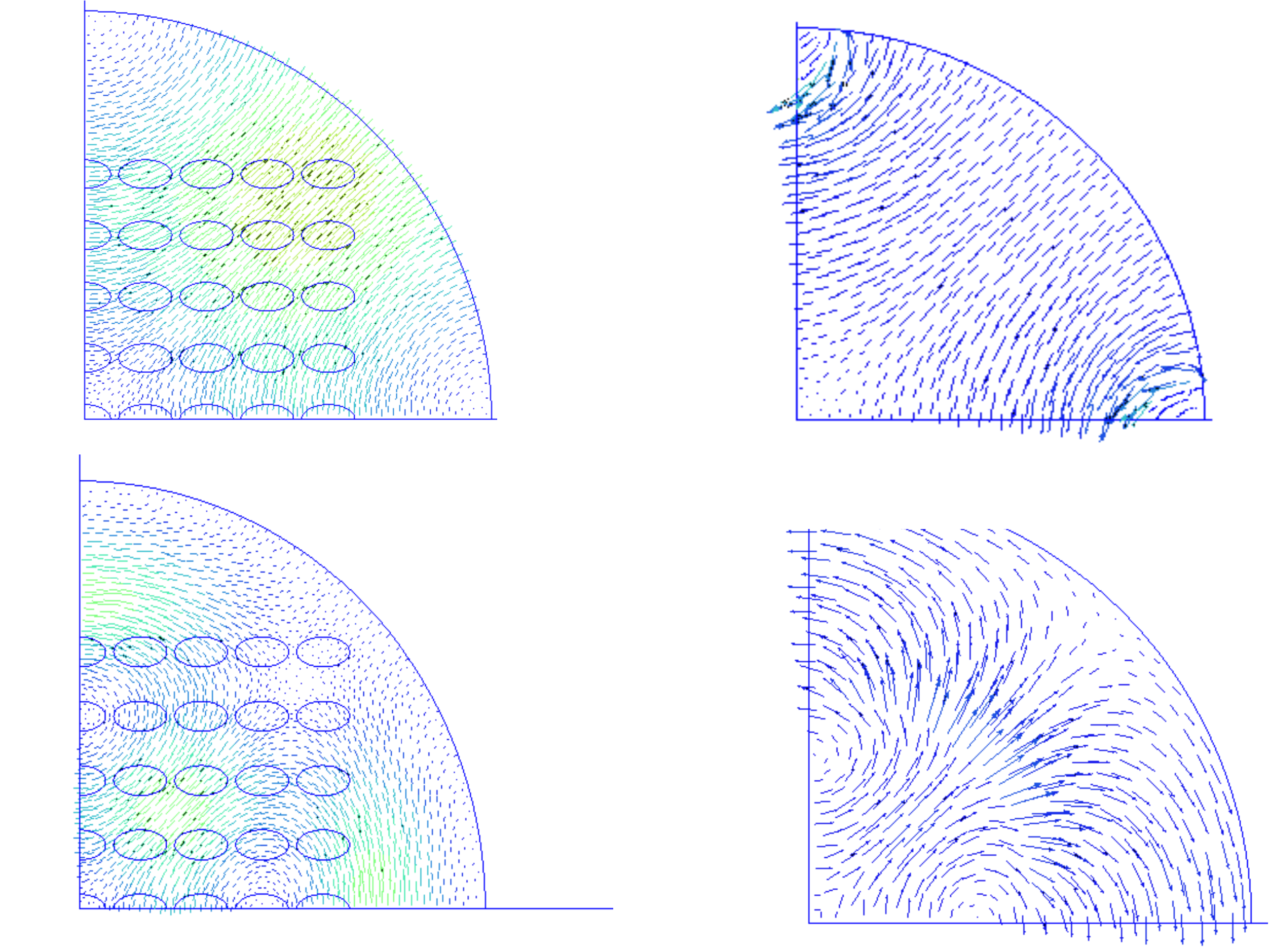} } } }
        \caption[anisotropy]{Comparison between transverse electric fields 
        $TE_{21}$ and $TE_{31}$ of a microstructured metallic waveguide
        for a propagation constant $\gamma=0.1 cm^{-1}$
        (wavenumbers $k=0.7707 cm^{-1}$ and $k=0.5478 cm^{-1}$
        respectively), see left panel, with the $TE_{21}$ and $TE_{31}$ modes of the corresponding
        homogenized anisotropic metallic waveguide for $\gamma = 0.1 cm^{-1}$ 
        ($k=0.7607 cm^{-1}$ and $k = 0.5201 cm^{-1}$, where
       $k=\sqrt{\omega^2/c^2-\gamma^2}=\sqrt{\omega^2\varepsilon_0\mu_0-\gamma^2}$
       were obtained from the computation of eigenvalues $\omega$
of homogenized problem $({\cal P}_0)$), see right panel.}
    \label{anisotropy}
\end{figure}

\subsection{The case of one-dimensional gratings: Application to invisibility cloaks}
\begin{figure}[h]\centerline{\scalebox{0.5}{\includegraphics{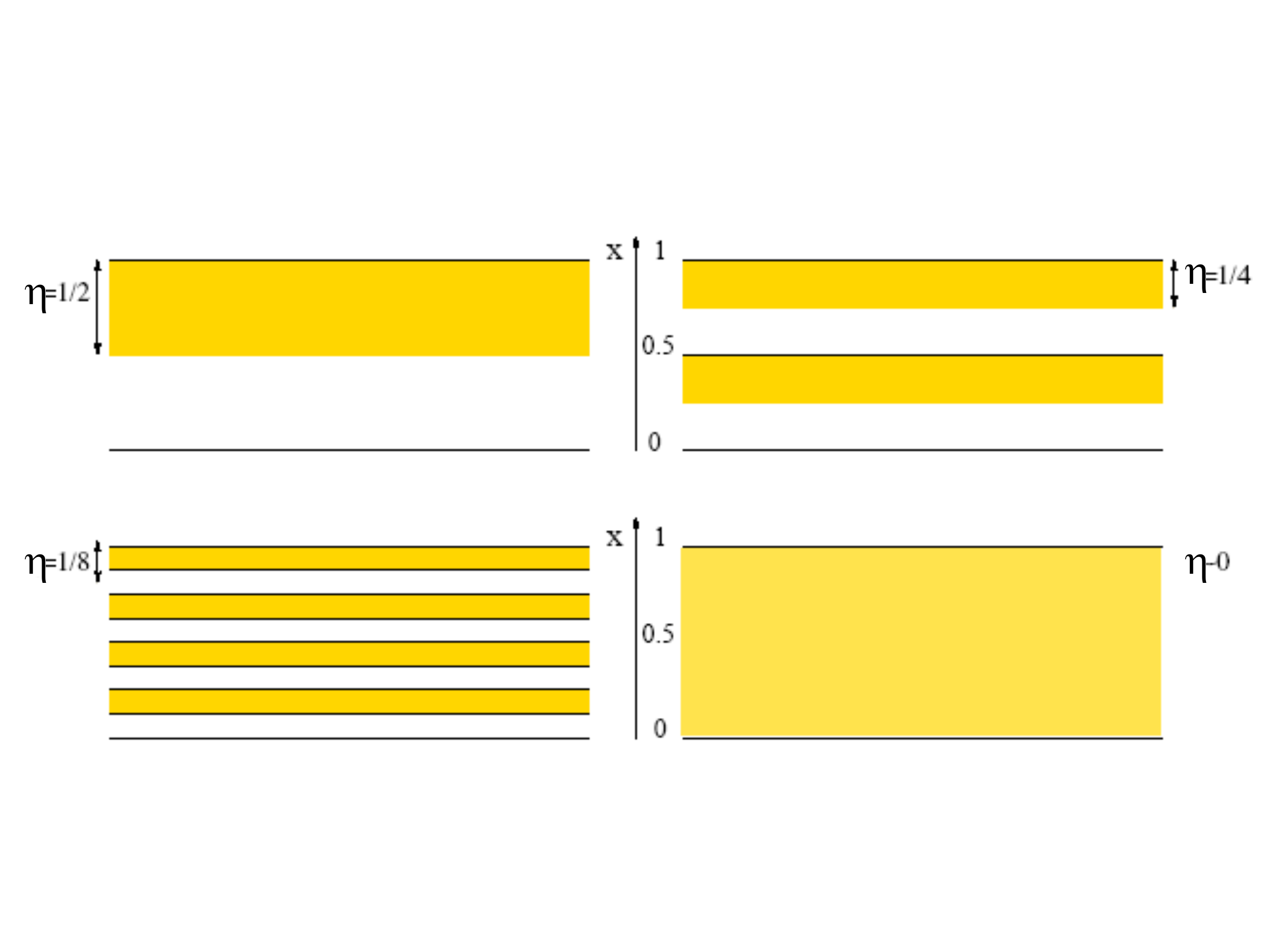}}}
\vspace{0cm}\caption{Schematic of homogenization process for a one-dimensional
grating with homogeneous dielectric layers of permittivity $\alpha$ and $\beta$
in white and yellow regions. When $\eta$ tends to zero the number of layers
tends to infinity, and their thicknesses vanish, in such a way that
the width of the overall stack remains constant.}
\label{fig2}
\end{figure}

There is a case of particular importance for applications in grating theory: that of a periodic
multilayered structure. Let us assume that the permittivity of this medium is
$\varepsilon=\alpha$ in white layers and $\beta$ in yellow layers,
as shown in Fig. \ref{fig2}.

\noindent Equation $({\cal A}2)$ takes the form:
$$({\cal A}3):-\frac{d}{dy}\left(\varepsilon^{-1}({y})\frac{d w}{\partial y}\right)
=\left(\frac{d}{dy}\varepsilon^{-1}({y})\right) \; ,$$
with $w({y})$, periodic function in $y$ of period $1$.

\noindent We deduce that
$$-\frac{d w}{d y}=1+C\varepsilon({y}) \; .$$

\noindent Noting that $\displaystyle{\int_Y \frac{d w}{dy}}=w(1)-w(0)=0$, this leads to
$$\int_Y \left( 1+C\varepsilon({y}) \right) dy = 0 \; .$$

\noindent Since $\mid Y \mid=1$, we conclude that
$$C=-{<\varepsilon>}^{-1} \; .$$

\noindent The homogenized permittivity takes the form:
\begin{equation}
\begin{array}{lll}
\varepsilon_{{\rm hom}}^{-1}&=\displaystyle{\frac{1}{\mid Y \mid}\int_Y\left( \varepsilon^{-1}({y})
+\varepsilon^{-1}({y})\frac{dw({y})}{dy}\right) \, dy} \nonumber \\
&= <\varepsilon^{-1}({y})> -<\varepsilon^{-1}({y})+C> \nonumber \\
&= <\varepsilon^{-1}({y})> -<\varepsilon^{-1}({y})>+<{<\varepsilon({y})>}^{-1}> = {<\varepsilon({y})>}^{-1} \; .  
\end{array}
\end{equation}

\noindent We note that if we now consider the full operator i.e. we include partial derivatives
in $y_1$ and $y_2$, the anisotropic homogenized permittivity takes the form:
$$
\varepsilon_{{\rm hom}}^{-1}=
\left(
\begin{array}{cc}
{<\varepsilon({y})^{-1}>} & 0 \\
0 & {<\varepsilon({y})>}^{-1}
\end{array}
\right) \; ,
$$
as the only contribution for $\varepsilon_{{\rm hom},11}^{-1}$ is $1/\mid Y\mid \int_Y \varepsilon^{-1}(y) \, dy$. 

\noindent As an illustrative example of what artificial anisotropy can achieve, we
propose the design of an invisibility cloak. For this, let us assume that we have a
multilayered grating with periodicity along the radial axis.
In the coordinate system $(r,\theta)$, the homogenized permittivity clearly has the same
form as above. If we want to design an invisibility cloak with an alternation of two
homogeneous isotropic layers of thicknesses $d_A$ and $d_B$ and
permittivities $\alpha$, $\beta$, we then need to use the formula
\begin{equation}
\begin{array}{lll}
&\displaystyle{\frac{1}{\varepsilon_r}}=\displaystyle{\frac{1}{1+\eta}\left(\frac{1}{\alpha}+\frac{\eta}{\beta}\right)},
&\varepsilon_\theta=\displaystyle{\frac{\alpha+\eta \beta}{1+\eta}} \; , \;
\nonumber
\end{array}
\label{effective}
\end{equation}
where $\eta=d_B/d_A$ is the ratio of thicknesses for layers $A$ and
$B$ and $d_A+d_B=1$.

\begin{figure}[h]
\hspace{-0.5cm}
\resizebox*{14cm}{!}{\includegraphics{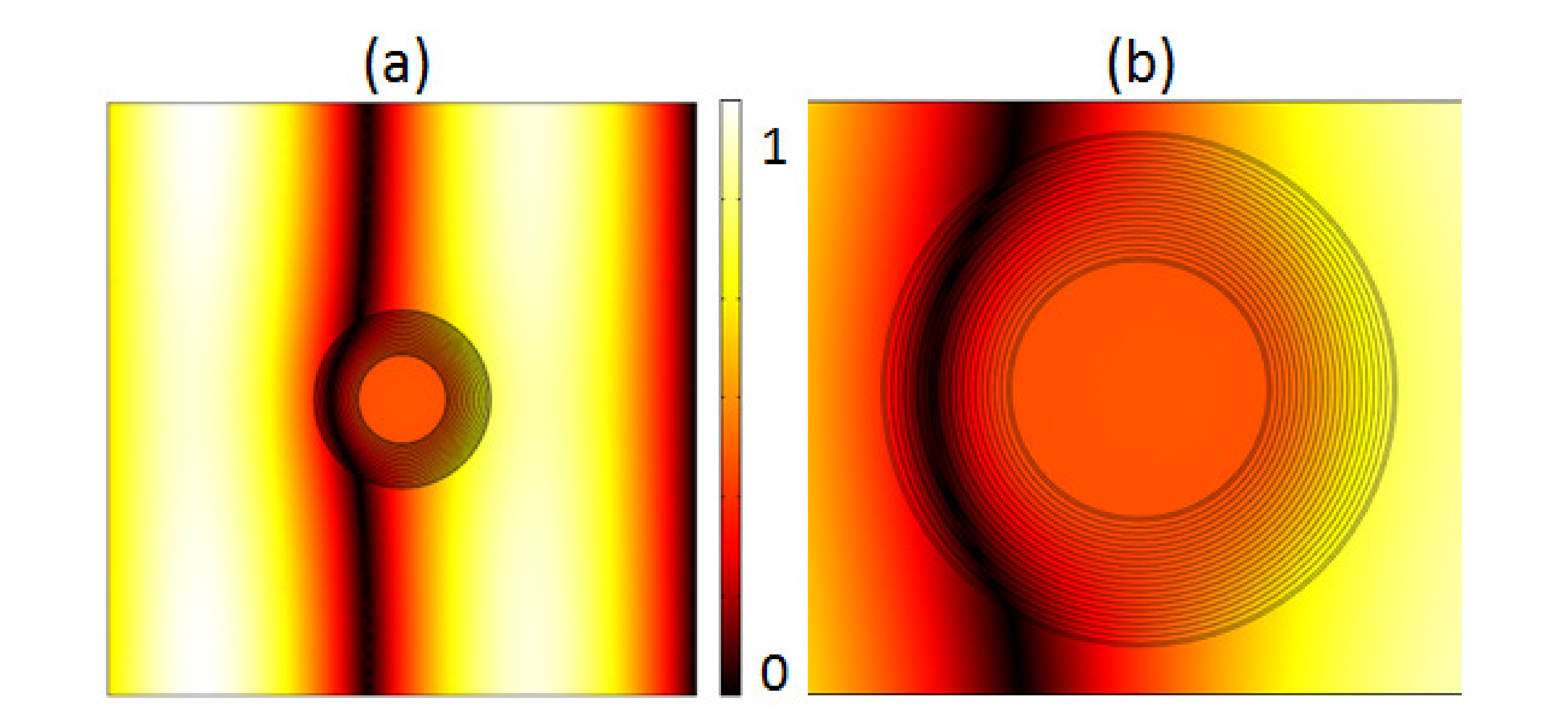}}
\vspace{0.0cm}\mbox{}
\caption{Propagation of a plane wave of wavelength $7 \; 10^{-7}$m
(red in the visible spectrum)
from the left on a multilayered cloak of inner radius $R_1=1.5 \; 10^{-8}$m
and outer radius $R_2=3 \; 10^{-8}$m,
consisting of 20 homogeneous layers of equal thickness and of respective
relative permittivities
$1680.70,0.25,80.75,0.25,29.39,0.25,16.37,0.25,10.99,0.25,8.18,0.25,6.50,0.25,5.40$,
$0.25,4.63,0.25,4.06,0.25$
in vacuum. Importantly, one layer in two has the same permittivity.
}
\label{lhf_fig1}
\end{figure}

We now note that the coordinate transformation
$r'=R_1+r\frac{R_2-R_1}{R_2}$ can compress a disc $r<R_2$ into
a shell $R_1<r<R_2$, provided that the shell is described by the
following anisotropic heterogeneous permittivity \cite{pendry}
$\underline{\underline{\varepsilon}}^{{\rm cloak}}$
(written in its diagonal basis):
\begin{equation}
\begin{array}{lll}
\varepsilon_r^{{\rm cloak}} &=\displaystyle{{\left(\frac{R_2}{R_2-R_1}\right)}^2}{\left(\frac{r'-R_1}{r'}\right)}^2 \; , \;
& \varepsilon_{\theta}^{{\rm cloak}} =\displaystyle{{\left(\frac{R_2}{R_2-R_1}\right)}^2} \; ,
\end{array}
\label{rhort1}
\end{equation}
where $R_1$ and $R_2$ are the interior and the exterior radii of the cloak.
Such a metamaterial can be approximated using the formula
(\ref{effective}), as first proposed in \cite{huang2007},
which leads to the multilayered cloak
shown in Fig. \ref{lhf_fig1}.

\section{High-frequency homogenization}\label{hfh}
Many of the features of interest in photonic crystals \cite{yablo,john}, or other
periodic structures, such as all-angle negative refraction \cite{zengerle87a,gralak2000,notomi2000,luo02a}
or ultrarefraction \cite{dowling,enoch2002} occur at high frequencies
where the wavelength and microstructure dimension are of similar orders.
Therefore the conventional low-frequency classical homogenisation clearly fails to
capture the essential physics and a different approach to distill the physics into an
effective model is required. Fortunately a high frequency homogenisation (HFH)
theory as developed in \cite{craster10a} is capable of capturing features such
as AANR and ultra-refraction \cite{josa2011} for some model structures.
Somewhat tangentially, there is an existing literature in the analysis community on
Bloch homogenisation \cite{allaire1998,allaire2005,birman,hoefer},
that is related to what we call
high frequency homogenisation. There is also a flourishing literature on developing
homogenised elastic media, with frequency dependent effective parameters, based
upon periodic media \cite{willis2011}. There is therefore considerable
interest in creating effective continuum models of microstructured media that break
free from the conventional low frequency homogenisation limitations.

\subsection{High Frequency Homogenization for Scalar Waves}
\label{sec:TMWaves}
Waves propagating through photonic crystals and metamaterials have proven to show different effects depending on their frequency. The homogenization of a periodic material is not unique. 
The effective
properties of a periodic medium change depending on the vibration
modes within its cells. The dispersion diagram structure can be considered to be the identity of such a material and provides the most important information regarding
group velocities, band-gaps of dis-allowed propagation frequency bands, Dirac cones and many other interesting effects. 
The goal of a homogenization theory is to provide an effective
homogeneous medium that is equivalent, in the long scale, to the
initial non-homogeneous medium composed of a short-scale periodic, or other
microscale, structure. This was achieved initially using the classical
theory of homogenization
\cite{Bensoussan78a,craster12a,jikov94a,mei96a,milton02a} and yields
an intuitively obvious result that the effective medium's properties
consist of simple averages of the original medium's
properties. This is valid so long as the wavelength is very large
compared to the size of the cells (here we focus on periodic media
created by repeating cells). For shorter wavelengths of the order of a
cell's length a more general theory has been developed
\cite{craster10a} that also recovers the results of the classical
homogenization theory. For clarity we present high frequency
homogeniaztion (HFH) by means of an illustrative example and consider
a two-dimensional lattice geometry for TE or TM polarised
electromagnetic waves. With harmonic time dependence, $\exp(-i\Omega
t)$ (assumed understood and henceforth suppressed),  the governing equation is the scalar Helmholtz equation,
\begin{equation}
\nabla^2 u+\Omega^2u=0,
\label{eq:Helmotz}
\end{equation}
where $u$ represent $E_Z$ and $H_Z$, for TM and TE polarised
electromagnetic waves respectively, and
$\Omega^2={n^2}\omega^2/{c^2}$. In our example the cells are square
and each square cell of length $2l$ contains a circular hole and the
filled part of the cell has constant non-dimensionalized
properties. The boundary conditions on the hole's surface, namely the boundary $\partial S_2$, depend on the polarisation and are taken to be either of Dirichlet or Neumann type. This approach assumes infinite conducting boundaries which is a good approximation for micro-waves. 
We adopt a multiscale approach where $l$ is the small length scale and
$L$ is a large length scale and we set $\eta = l/L\ll 1$ to be the
ratio of these scales. The two length scales let us introduce the
following two independent spatial variables, $\xi_i = x_i/l$ and $X_i = x_i/L$. The cell's reference coordinate system is then $-1<\xi<1$. By introducing the new variables in equation (\ref{eq:Helmotz}) we obtain,
\begin{equation}
u({\bf X},\bxi),_{\xi_i \xi_i}+\Omega^2 u({\bf X},\bxi)+ 2\eta  u({\bf X},\bxi),_{\xi_i X_i}+\eta^2u({\bf X},\bxi),_{X_iX_i}=0.
\label{eq:NewEquation}
\end{equation}
We now pose an ansatz for the field and the frequency, 
\begin{align}
u({\bf X},\bxi)=u_0({\bf X},\bxi)+\eta u_1({\bf X},\bxi)+\eta^2 u_2({\bf X},\bxi)+\ldots , 
\nonumber\\ 
\Omega^2=\Omega_0^2+\eta \Omega_1^2+\eta^2 \Omega_2^2+\ldots
\label{eq:expansion2D}
\end{align}
In this expansion we set $\Omega_0$ to be the frequency of standing
waves that occur in the perfectly periodic setting. 
By substituting equations (\ref{eq:expansion2D}) into equation
(\ref{eq:NewEquation}) and grouping equal powers of $\epsilon$ through
to  second order, we obtain a hierarchy of three ordered equations:
\begin{equation}
u_{0,\xi_i\xi_i} + \Omega_0^2u_0=0,
\label{eq:leadingOrder}
\end{equation}
\begin{equation}
u_{1,\xi_i\xi_i} + \Omega_0^2u_1=
-2u_{0,\xi_iX_i}
-\Omega_1^2 u_0,
\label{eq:firstOrder}
\end{equation}
\begin{equation}
 u_{2,\xi_i\xi_i} + \Omega_0^2 u_2 =-u_{0,X_iX_i}
   -2u_{1,\xi_iX_i}
-\Omega_1^2 u_1 -\Omega_2^2 u_0.
  \label{eq:secondOrder}
\end{equation}
These equations are solved as in \cite{antonakakis13a,craster10a} and
hence the description is brief. 

\begin{figure}
  \begin{center}
    \includegraphics[scale=0.5]{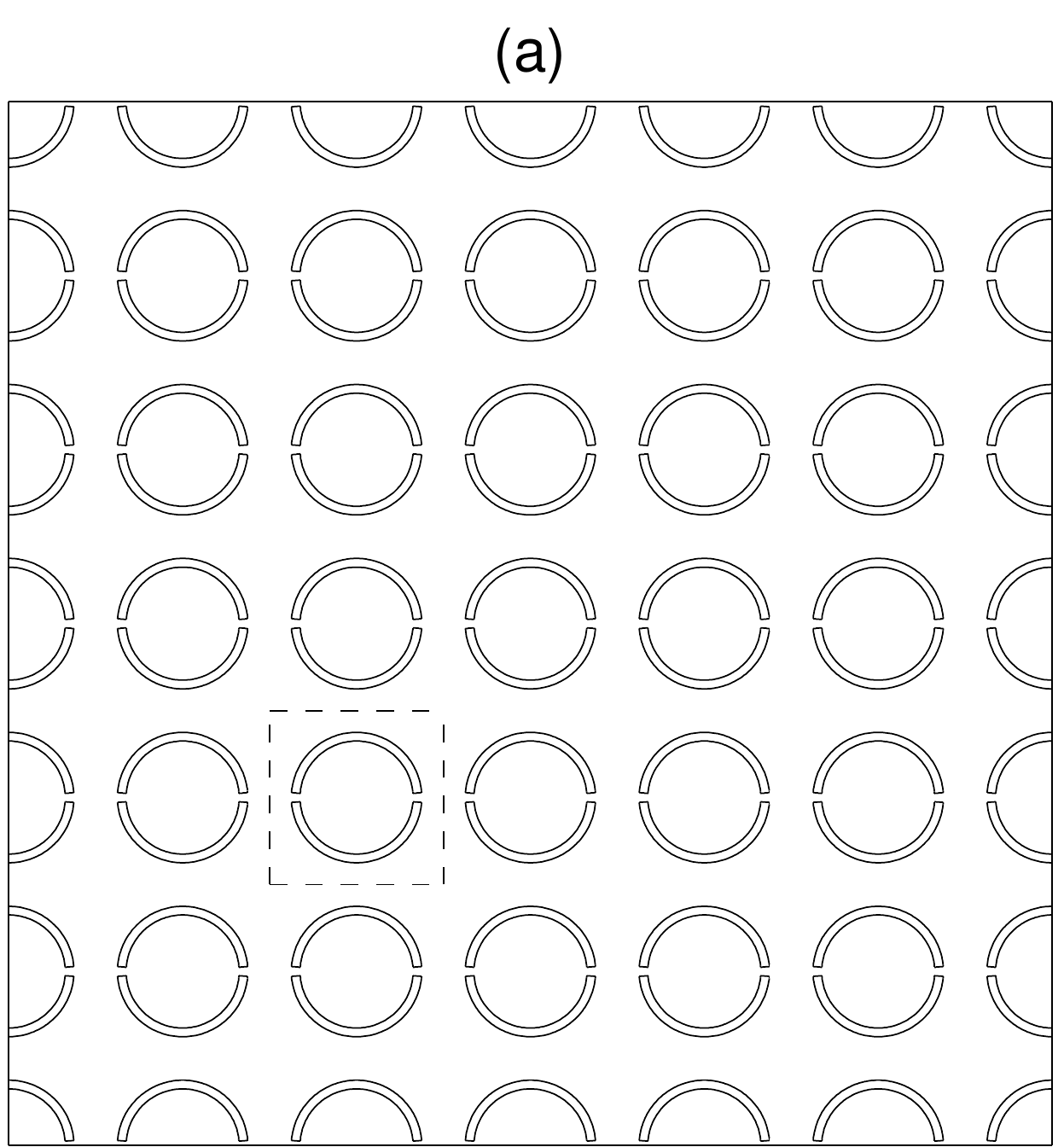}
  \includegraphics[scale=0.9]{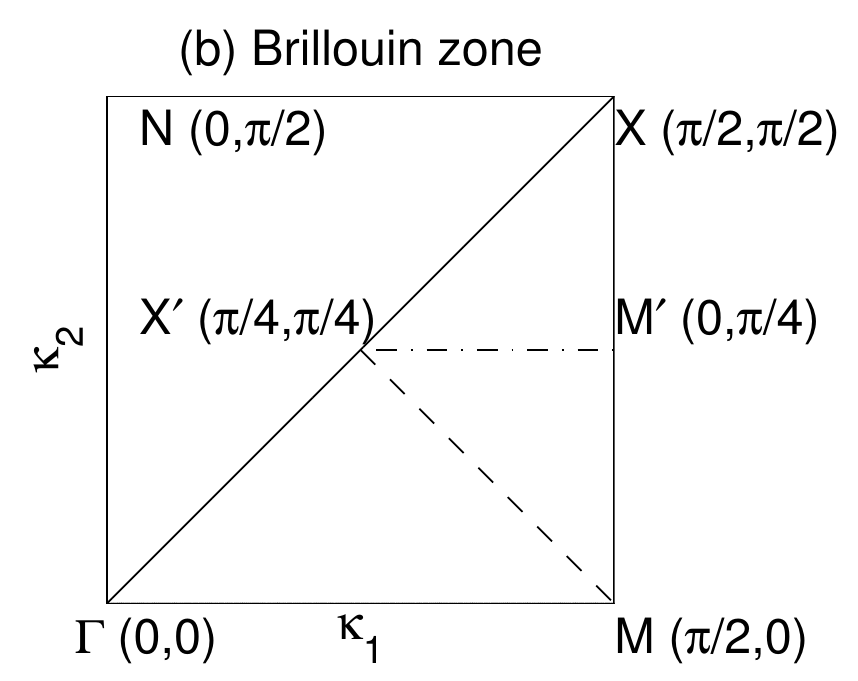}
   \end{center}
\caption{Panel (a) An infinite square array of split ring resonators with the
  elementary cell shown as the dashed line inner square. Panel (b)
  shows the irreducible Brillouin zone, in wavenumber space, used for
  square arrays in perfectly periodic
  media based around the elementary cell shown of length $2l$ ($l=1$ in (b)).
Figure reproduced from Proceedings of the Royal Society \cite{antonakakis13a}.}
\label{fig:Brillouin}
\end{figure}

The asymptotic expansions are taken about the standing wave
frequencies that occur at the corners of the irreducible Brillouin
zone depicted in Fig. \ref{fig:Brillouin}. It should be noted that not
all structured cells will have the usual symmetries of a square, as in
Fig. \ref{fig:Brillouin}(a) where there is no reflexion symmetry from
the diagonals. As a consequence the usual triangular region $\Gamma XM$
does not always represent the irreducible Brillouin zone and the
square region $\Gamma MXN$ should be used instead. Also paths that
cross the irreducible Brillouin zone have proven to yield interesting
effects namely along the path $MX'$ for large circular holes
\cite{craster12b}. 

The subsequent asymptotic development considers small perturbations about the points $\Gamma$, $X$ and $M$ so that the boundary conditions of $u$ on the outer boundaries of the cell, namely $\partial S_1$, read,
\begin{equation}
u|_{\xi_{i}=1}=\pm u|_{\xi_{i}=-1} \quad \text{and} \quad u_{,\xi_i}|_{\xi_{i}=1}=\pm u_{,\xi_i}|_{\xi_{i}=-1},
\label{eq:periodicBC}
\end{equation}
where the $+,-$ stand for periodic and anti-periodic conditions
respectively: the standing waves occur when these conditions are met. 
The conditions on $\partial S_2$ are either of Dirichlet or Neumann type. The theory that follows is similar for both boundary condition cases, but the latter one is illustrated herein. Neumann boudary condition on the hole's surface or equivalently electromagnetic waves in TE polarization yield,
\beq
\frac {\partial u} {\partial {\bf n}} =u_{,x_i} n_i|_{\partial S_2}=0.
\label{eq:neumann}
\eeq
which in terms of the two-scales and $u_i({\bf X}, \bxi)$ become
\beq
U_{0,\xi_i}n_i = 0,\quad
\label{eq:1rstNeumann}
(U_0f_{0,X_i} + u_{1,\xi_i})n_i=0,\quad
( u_{1,X_i} + u_{2,\xi_i}) n_i=0.
\eeq   
The solution of the leading order equation is by 
introducing the following separation of variables $u_0=f_0({\bf
  X})U_0(\bxi;\Omega_0)$. It is obvious that $f_0({\bf X})$, which
represents the behaviour of the solution in the long scale, is not set
by the leading order equation and the resulting eigenvalue problem is
solved on the short-scale for $\Omega_0$ and $U_0$ representing the
standing wave frequencies and the associated cell's vibration modes
respectively. 
To solve the first order equation (\ref{eq:firstOrder}) we  take the
integral over the cell of the product of equation
(\ref{eq:firstOrder}) with $U_0$ minus the product of equation
(\ref{eq:leadingOrder}) with $u_1/f_0$ and this yields
$\Omega_1=0$. It then follows to solve for $u_1({\bf
  X},\bxi)=f_{0,X_i}({\bf X})U_{1_i}(\bxi)$ where the vector ${\bf
  U}_1$ is found as in \cite{antonakakis13a}. By invoking a similar
solvability condition for the second order equation we obtain a second
order PDE for $f_0({\bf X})$, 
\begin{align}
T_{ij}f_{0,X_iX_j}+\Omega_2^2f_0=0\quad{\rm where},
\nonumber\\
\quad T_{ij}=\frac{t_{ij}}{\dint_S U_0^2dS}\quad{\rm for} \quad i,j=1,2
\label{eq:f_0}
\end{align}
 entirely on the long scale with the coefficients $T_{ij}$ containing all
 the information of the cell's dynamical response and 
 the tensor $t_{ij}$ represents dynamical averages of the properties of the medium. For Neumann boundary conditions on $\partial S_2$ its formulation reads,
\beq
t_{ii}=\dint_SU_0^2dS+\dint_S(U_{1_i,\xi_i}U_0-U_{1_i}U_{0,\xi_i})dS
\quad {\rm for} \quad i=1\ {\rm or}\ 2,
\label{eq:t11}
\eeq
\beq
t_{ij}=\dint_S(U_{1_j,\xi_i}U_0-U_{1_j}U_{0,\xi_i})dS \quad {\rm for} \quad i\neq j.
\label{eq:tij}
\eeq
Note that there is no summation over repeated indexes for $t_{ii}$.
The tensor depends on the boundary conditions of the holes and has a
different form if Dirichlet type conditions are applied on $\partial
S_2$.

The PDE for $f_0$ has several uses, and can be verified by re-creating
 asymptotically the dispersion curves for a perfect lattice system.  
One important result of equation (\ref{eq:f_0}) is its use in the expansion of $\Omega$ namely in equation (\ref{eq:expansion2D}). In order to obtain $\Omega_2$ as a function of the Bloch wavenumbers we use the Bloch boundary conditions on the cell to solve for $f_0({\bf X})=\exp(i\kappa_jX_j/\eta)$, where $\kappa_j=K_j-d_j$ with $d_j=0,\pi/2,-\pi/2$ depending on the location in the Brillouin zone. The asymptotic dispersion relation now reads,
\beq
\Omega\sim\Omega_0+ \frac{T_{ij}}{2\Omega_0}\kappa_i \kappa_j.
\label{eq:asymptoticDispersion}
\eeq
Equation (\ref{eq:asymptoticDispersion}) yields the behaviour of the dispersion curves asymptotically around the standing wave frequencies that are naturally located at the edge points of the Brillouin zone. Fig. \ref{fig:2D_Neumann} illustrates the asymptotic dispersion curves for the first six dispersion bands of a square cell geometry with circular holes.
\begin{figure}
  \begin{center}
    \includegraphics[scale=0.8]{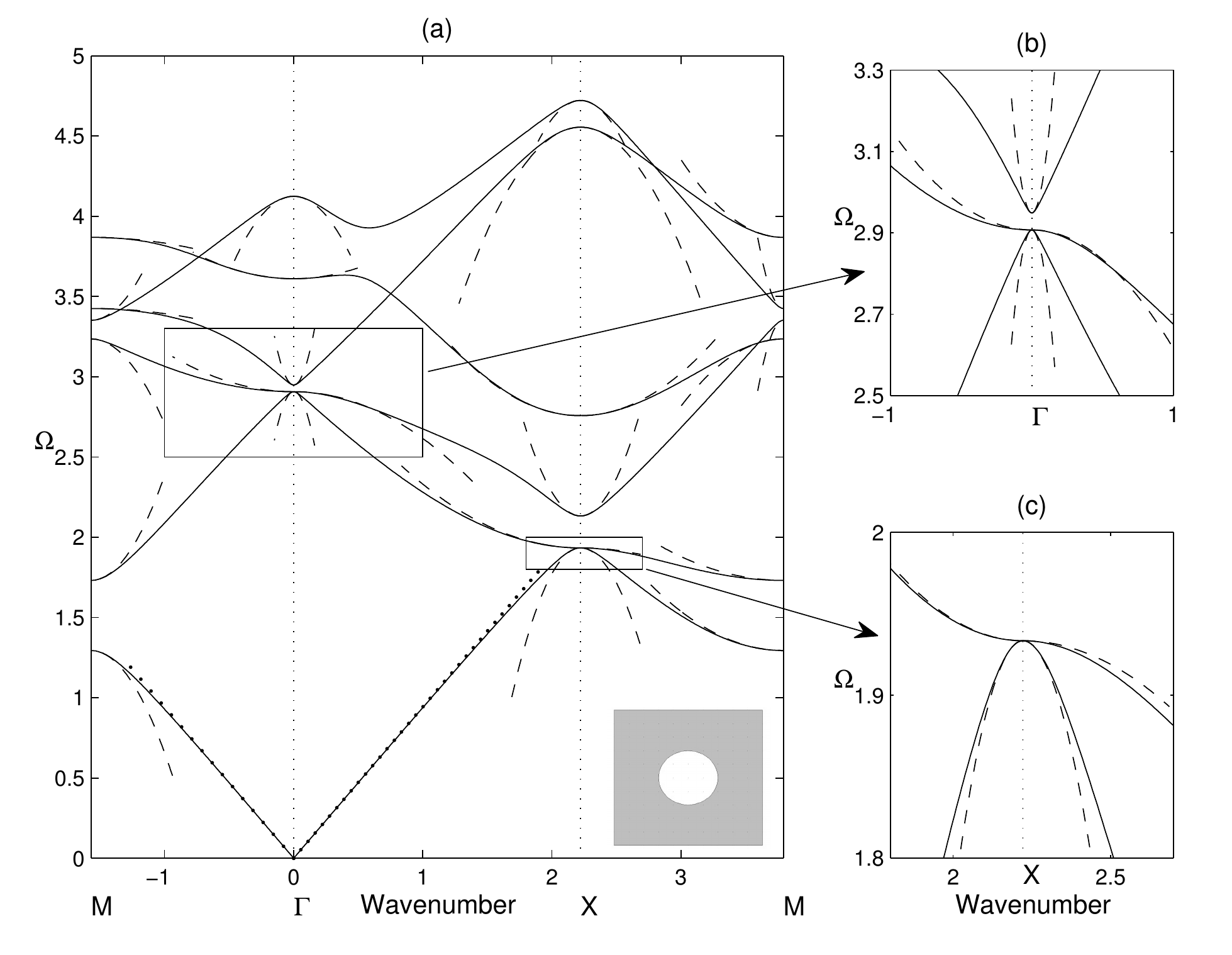}
   \end{center}
\vspace{-1cm}
\caption{The dispersion diagram for a doubly periodic array of square cells with circular inclusions, of radius $0.4$, free at their inner boundaries shown for the irreducible Brillouin zone of
  Fig. \ref{fig:Brillouin}. The dispersion curves are shown in solid lines and the asymptotic solutions from the high frequency homogenization theory are shown in dashed lines.
Figure reproduced from Proceedings of the Royal Society \cite{antonakakis13a}.}
\label{fig:2D_Dirichlet}
\end{figure}

An assumption in the development of equation
(\ref{eq:asymptoticDispersion}) is that the standing wave frequencies
are isolated. But one can clearly see in Fig. \ref{fig:2D_Dirichlet}
that this is not the case for third standing wave frequency at point
$\Gamma$ as well as for the second standing wave frequency at point
$X$. A small alteration to the theory \cite{antonakakis13a} enables the computation of the dispersion curves at such points by setting,
\beq
u_0=f_0^{(l)}({\bf X})U_0^{(l)}(\bxi;\Omega_0)
\label{eq:uzeroRep}
\eeq
 where we sum over the repeated superscripts $(l)$. 
Proceeding as before, we multiply equation (\ref{eq:firstOrder}) by $U_0^{(m)}$, substract $u_1((U^{(m)}_{0,\xi_i})_{\xi_i}+\Omega_0^2 U_0^{(m)})$ then integrate over the cell to obtain,
\beq
\left(\frac{\partial}{\partial X_j}{\bf A}_{jml}+\Omega_1^2{\bf B}_{ml}\right){\hat f}_0^{(l)}=0,\quad{\rm for}\quad m=1,2,\ldots,p
\label{eq:PreSystem}
\eeq
  $\Omega_1$ is not necessarily zero, 
and 
\beq
{\bf
  A}_{jml}=\dint_S(U_0^{(m)}U_{0,\xi_j}^{(l)}-U_{0,\xi_j}^{(m)}U_0^{(l)})dS, \quad
{\bf B}_{ml}=\dint U_0^{(l)}U_0^{(m)}dS.
\label{eq:Bmatrix}
\eeq

 There is now a system of coupled partial differential equations for
 the $f_0^{(l)}$ and, provided $\Omega_1\neq 0$, the leading order
 behaviour of the dispersion curves near the $\Omega_0$ is now linear
 (these then form Dirac cones).  

For the perfect lattice, we set $f_0^{(l)}={\hat
  f}_0^{(l)}\exp(i\kappa_jX_j/\eta)$ and 
 obtain the following index equations,
\beq
(i\frac{\kappa_j}{\eta}{\bf A}_{jml}+\Omega_1^2{\bf B}_{ml}){\hat f}_0^{(l)}=0,  \quad \rm{for} \quad m=1,2,...,p
\label{eq:PreSystem2}
\eeq
The system of equation (\ref{eq:PreSystem2}) can be written simply as,
\beq
{\bf C}{\hat {\bf F}}_0=0,
\label{eq:System}
\eeq
with ${\bf C}_{ll}=\Omega_1^2{\bf B}_{ll}$ and ${\bf
  C}_{ml}=i\kappa_j{\bf A}_{jml}/\eta$ for $l\neq m$. One must then solve
for $\Omega_1^2=\pm\sqrt{\alpha_{ij}\kappa_i\kappa_j}/\eta$ when the determinant of ${\bf C}$ vanishes and insert the result in,
\beq
\Omega\sim\Omega_0\pm\frac{1}{2\Omega_0}\sqrt{\alpha_{ij}\kappa_i\kappa_j}.
\label{eq:asymptoticExpansionLinear}
\eeq
If the $\Omega_1$ are zero one must go to the next order.

\subsubsection{Repeated eigenvalues: quadratic asymptotics}
\label{sec:RepEigenQuad}
If $\Omega_1$ is zero, $u_1=f_{0,X_k}^{(l)}U_{1_k}^{(l)}$ (we again
sum over all repeated $(l)$ superscripts) and 
we advance to second order using
(\ref{eq:secondOrder}). Taking the difference between
the product of equation (\ref{eq:secondOrder}) with $U_0^{(m)}$ and
$u_2(U_{0,\xi_i\xi_i} + \Omega_0^2 U_0)$ and then
integrating 
over the elementary cell gives
 \ba
&f_{0,X_iX_i}^{(l)}\dint_SU_0^{(m)}U_0^{(l)}dS+
f_{0,X_kX_j}^{(l)}\dint_S(U_0^{(m)}U_{1_k,\xi_j}^{(l)}-U_{0,\xi_j}^{(m)}U_{1_k}^{(l)})dS
\nonumber\\
&\quad +\Omega_2^2 f_0^{(l)}\dint_S U_0^{(m)}U_0^{(l)}dS=0, \quad
{\rm for} \quad m=1,2,...,p
\label{eq:preLam2}
\ea
 as a system of coupled PDEs. 
The above equation is presented more neatly as 
\beq
f_{0,X_iX_i}^{(l)}{\bf A}_{ml}+f_{0,X_kX_j}^{(l)}{\bf
  D}_{kjml}+\Omega_2^2 f_0{\bf B}_{ml}=0, \quad {\rm for} \quad m=1,2,...,p.
\label{eq:Lam2}
\eeq
For the Bloch wave setting, using $f_0^{(l)}({\bf X})={\hat f}_0^{(l)}\exp(i\kappa_jX_j/\eta)$ we obtain the following system,
\beq
\left(-\frac{\kappa_i\kappa_i}{\eta^2}{\bf A}_{ml}-\frac{\kappa_k\kappa_j}{\eta^2}{\bf
  D}_{kjml}+\Omega_2^2{\bf B}_{ml}\right){\hat f}_0^{(l)}=0, \quad {\rm for} \quad m=1,2,...,p
\label{eq:sys3}
\eeq
 and this determines the asymptotic dispersion curves.

\subsubsection{The classical long wave zero frequency limit}
\label{sec:classical}
The current theory simplifies if one enters the classical long wave,
low frequency limit where $\Omega^2\sim O(\epsilon^2)$
 as $U_0$ becomes uniform, and without loss of generality is set to be
 unity, over the elementary cell. The final equation is again
 (\ref{eq:f_0}) where the tensor $t_{ij}$ simplifies to 
\beq
t_{ii}=\dint_SdS+\dint_SU_{1_i,\xi_i}dS,\quad 
t_{ij}=\dint_SU_{1_j,\xi_i}dS \quad {\rm for} \quad i\neq j
\label{eq:tij0}
\eeq
 (with no summation over  repeated suffices in this equation)
and $T_{ij}=t_{ij}/\dint_S dS$.

\subsection{Illustrations for Tranverse Electric Polarized Waves}
\label{sec:EMWaves}
\begin{figure}
  \begin{center}
    \includegraphics[scale=0.7]{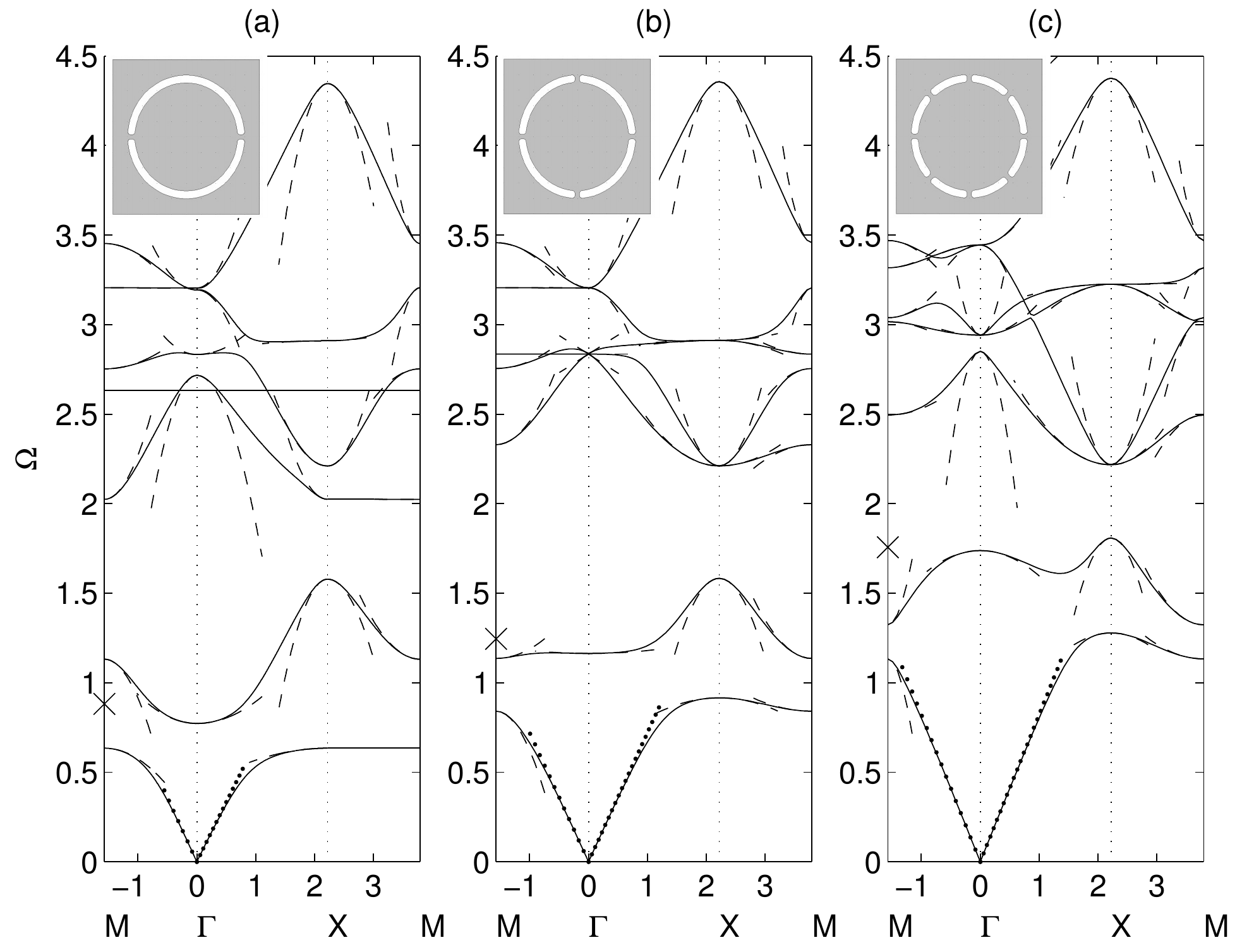}
   \end{center}
\vspace{-1.0cm}
\caption{The dispersion diagrams for a doubly periodic array of square cells with split ring inclusions, free at their inner boundaries shown for the irreducible Brillouin zone of
  Fig. \ref{fig:Brillouin}. The dispersion curves are shown in solid lines and the asymptotic solutions from the high frequency homogenization theory are shown in dashed lines.
Figure reproduced from Proceedings of the Royal Society \cite{antonakakis13a}.}
\label{fig:2D_Neumann}
\end{figure}
 Let us now turn to some illustrative examples. We present in
 Fig. \ref{fig:2D_Neumann} the TE polarization waves for three types
 of SRR's (Split Ring Resonator's). Equation (\ref{eq:f_0}) represents
 the wave propagation in the effective medium. It is noticable that
 the $T_{ij}$ coefficients depend on the standing wave frequency and
 that $T_{11}$ is not necessarily equal to $T_{22}$ in order to yield
 an anisotropic effective medium for each separate frequency. Near
 some of the standing wave frequencies the anisotropy effects are very
 pronounced and well explained by the no longer elliptic equation
 (\ref{eq:f_0}).

In the above equations $U_{1_i}$ is a solution of,
\beq
U_{1_j,\xi_i\xi_i} =0,
\label{eq:firstOrderHomo}
\eeq
with boundary conditions $(f_{0,X_i} + u_{1,\xi_i})n_i=0$ on
the hole boundary. If the medium is homogeneous as it is in the illustrative examples herein, equation (\ref{eq:firstOrderHomo}) is the same as
that for $U_0$, but with different
boundary conditions. The specific boundary conditions for $U_{1_j}$ are
\beq
U_{1_j,\xi_i}n_i=-n_j\quad\text{for}\quad j=1,2,
\label{eq:NeumannU1}
\eeq
where $n_i$ represent the normal vector components to the hole's surface. The role of ${\bf
  U_1}$ is to ensure Neumann boundary conditions hold and the tensor
contains simple averages of inverse permittivity and permeability supplemented by
the correction term which takes into account the boundary conditions at $\partial S_2$.
 Equation 
(\ref{eq:tij0}) is the classical expression for the homogenised coefficient
in a scalar wave equation with constant material properties; (\ref{eq:firstOrderHomo})
is the well-known annex problem of electrostatic type set on a
periodic cell, see 
\cite{Bensoussan78a,jikov94a}, and also holds
for the homogenised vector Maxwell's system, where ${\bf U_1}$ now has three components
and $i,j=1,2,3$ \cite{zolla00a,wellander,zolla07a}. 

\subsubsection{Cloaking in metamaterials}

\begin{figure}
\vspace{-0.5cm}
  \begin{center}
    \includegraphics[scale=0.15]{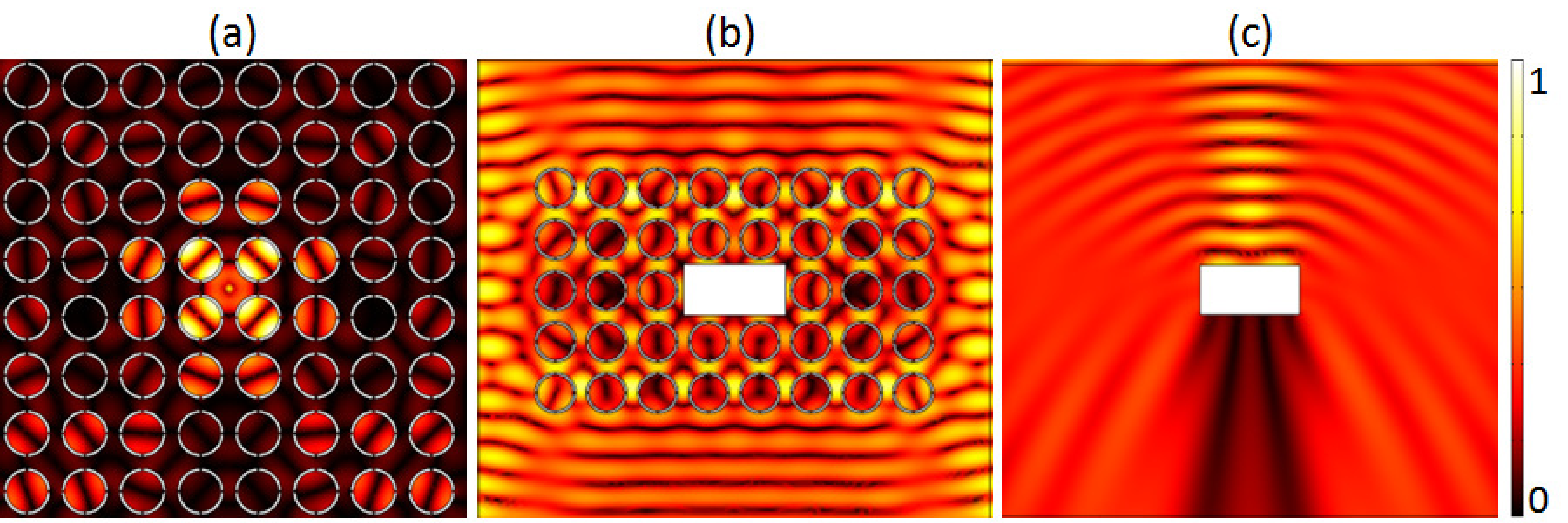}
 \includegraphics[scale=0.6]{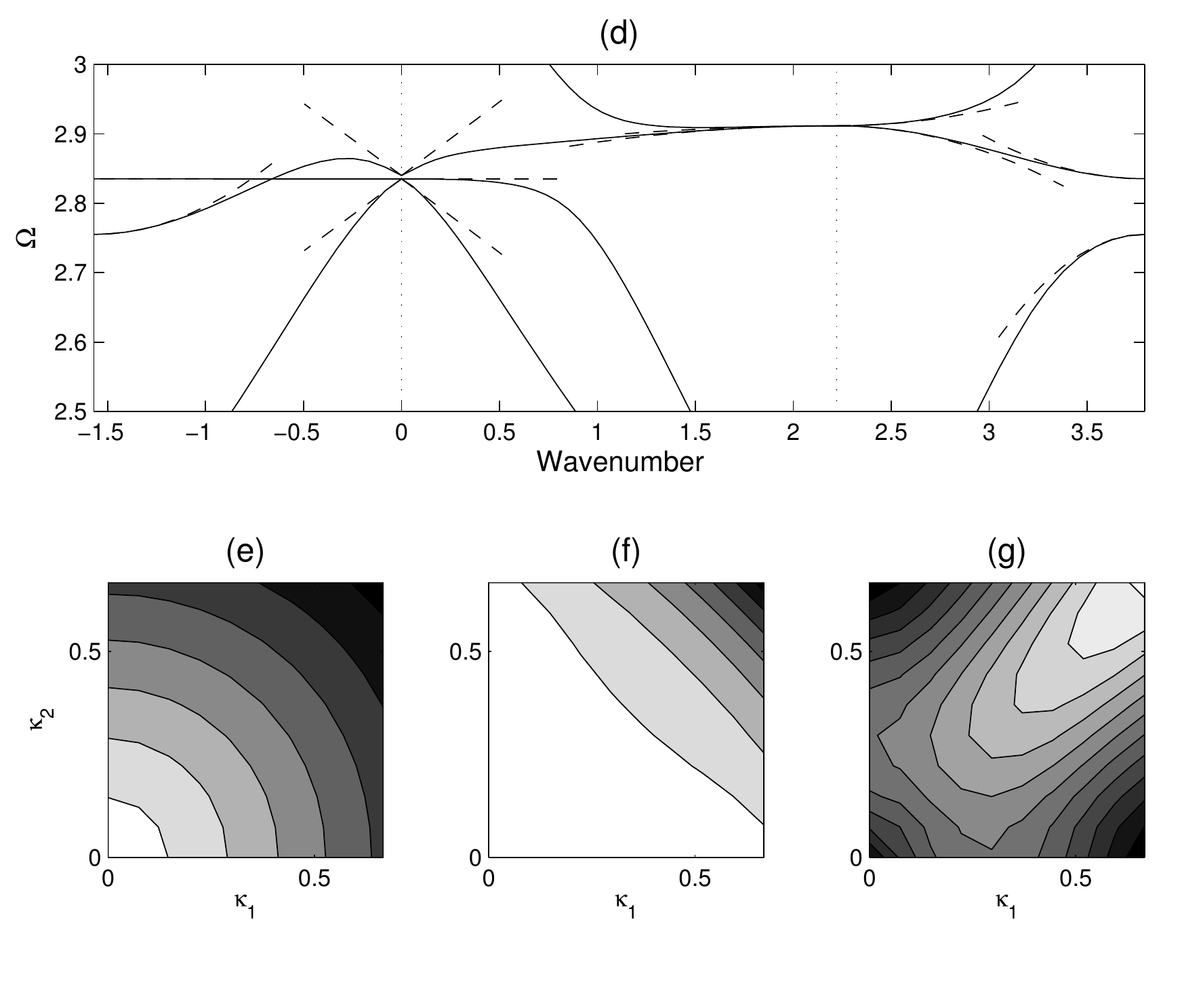}
   \end{center}
\vspace{-1.4cm}
\caption{Cloaking in square arrays of SRRs with four holes:
A source at frequency $\Omega=2.8$, located in the center of a square metamaterial
consisting of 64 SRRs shaped as in Fig. \ref{fig:2D_Neumann}(b) produces a wave pattern
reminiscent of (a)  concentric spherical field, (b) cloaking of a rectangular inclusion inside a slab of a metamaterial consisting of 38 SRRs
and (c) scattering of a plane wave from the same rectangular hole as the previous panel. (d) Zoom in dispersion diagram of
Fig. \ref{fig:2D_Neumann}(b). Panels (e), (f) and (g) present isofrequency plots of the respective the lower, middle and upper modes of the Dirac point. Figure reproduced from Proceedings of the Royal Society \cite{antonakakis13a}.}
\label{fig:unionjack}
\end{figure}

 SRRs with 4 holes are now used and the dispersion diagrams are in 
Fig. \ref{fig:2D_Neumann} (b). The flat band along the $M\Gamma$ path
is interesting for the fifth mode and we choose to illustrate cloaking
effects that occur here. In Fig. \ref{fig:unionjack}(a), we set an harmonic source at the
corresponding frequency $\Omega=2.8$ in an $8\times 8$ array of SRRs and
observe a wave pattern of concentric spherical modes. As can be seen in Figs. \ref{fig:unionjack}(b) and \ref{fig:unionjack}(c) a plane wave propagating at frequency $\Omega=2.8$ demonstrates perfect transmission through a slab composed of 38 SRRs but also cloaking of a rectangular inclusion where no scattering is seen before or after the metamaterial slab. Panel (d) of Fig. \ref{fig:unionjack} shows the location in the band structure that is responsible for this effect. Note that the frequency of excitation is just below the Dirac cone point located at $\Omega=2.835$ where the group velocity is negative but also constant near that location of the Brillouin zone illustrated through an isofrequency plot of lower mode of the Dirac point in Fig. \ref{fig:unionjack}(e). 
 In constrast with the isotropic features of panel (e), those of panels (f) and (g) show 
ultra-flattened isofrequency contours that relate to
ultra-refraction, a regime more prone to omni-directivity than
cloaking. The asymptotic system of equations (\ref{eq:PreSystem}) describing the effective medium at the Dirac point can be uncoupled to yield one same equation for all $f_0^{(j)}$'s,
\beq
f_{0,X_iX_i}^{(j)}+0.7191\Omega_1^4 f_0^{(3)}=0
\label{eq:f03}
\eeq

After some further analysis, the PDE for $f_0^{(2)}$ is responsible for the effects at the frequency chosen $\Omega=2.8$.

\subsubsection{Lensing via AANR and St Andrew's cross in metamaterials}

\begin{figure}
\vspace{-0.5cm}
  \begin{center}
\includegraphics[scale=0.15]{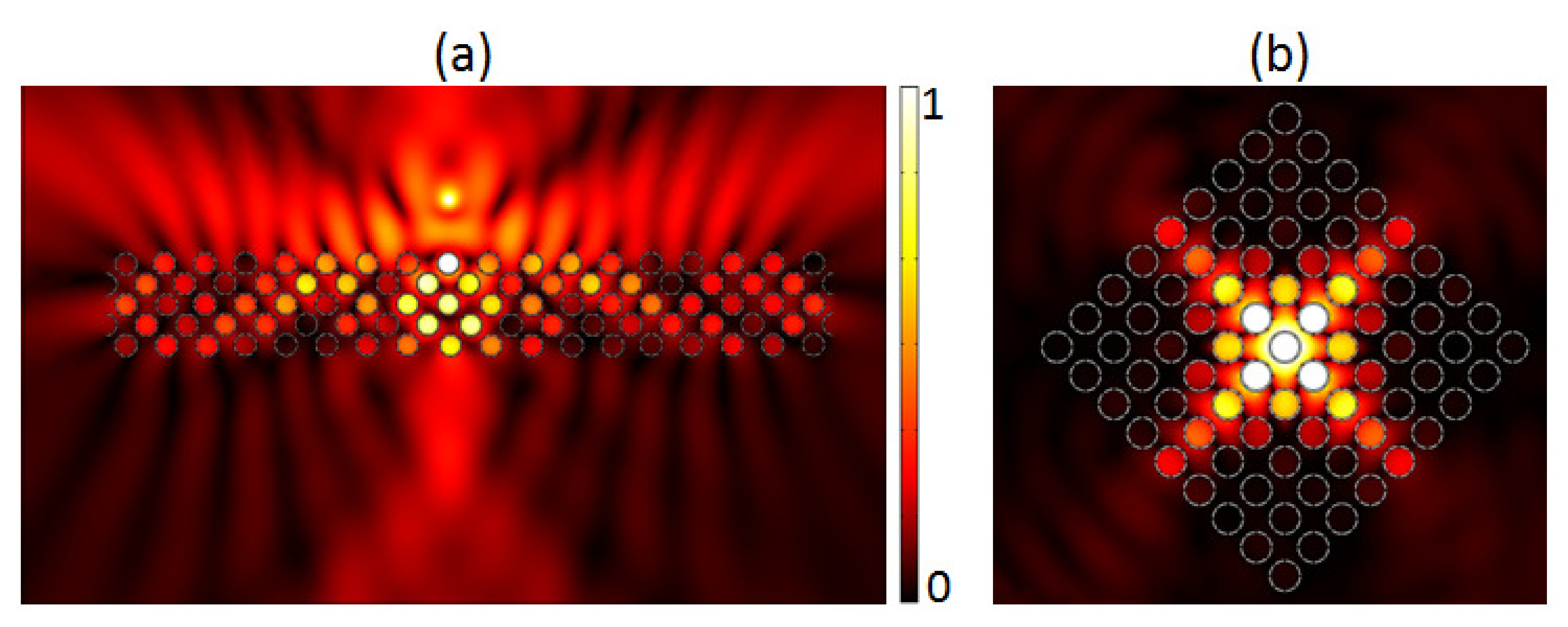}
 \includegraphics[scale=0.6]{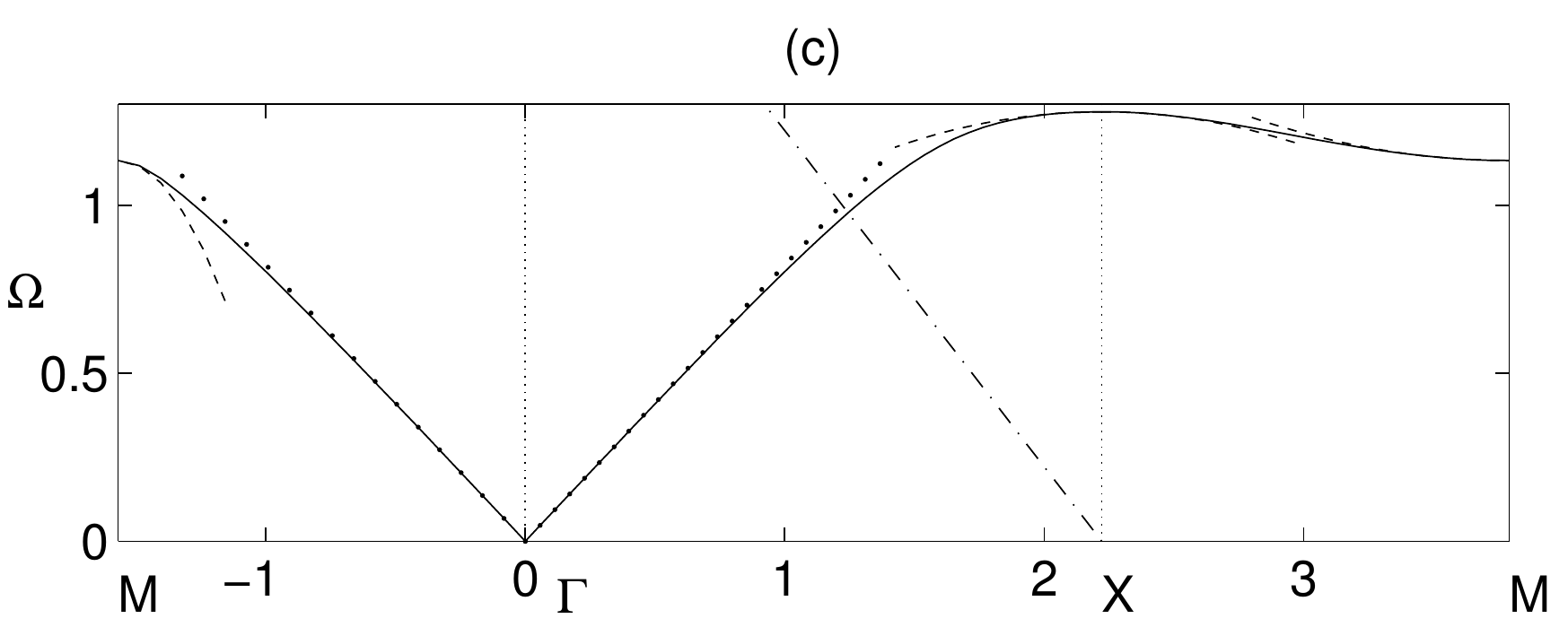}
\end{center}
\vspace{-1cm}
\caption{Lensing via AANR and St Andrew's cross in square arrays of SRRs with eight holes:
(a) A line source at frequency $\Omega=1.1375$ located above a rectangular
metamaterial consisting of of 90 SRRs as in Fig. \ref{fig:2D_Neumann}(c) displays an image underneath
(lensing);
(b) A line source at frequency $\Omega=1.25$ located inside a square
metamaterial consisting of 49 SRRs as in Fig. \ref{fig:2D_Neumann}(c) displays the dynamically induced anisotropy of the effective medium;
(c) Zoom in dispersion diagram of Fig. \ref{fig:2D_Neumann}(c).
Note that each cell in the arrays in (a) and (b) has been rotated through an angle $\pi/4$.
Figure reproduced from Proceedings of the Royal Society \cite{antonakakis13a}.}
\label{fig:aanrbis}
\end{figure}

We observe all-angle-negative-refraction effect in
metamaterials with SRRs with 8 holes. The dispersion curves in
Fig. \ref{fig:2D_Neumann}(c) are interesting, as the second curve
displays the hallmark of an optical band for a photonic crystal (it
has a negative group velocity around the $\Gamma$ point).  However,
this band is the upper edge of a low frequency stop band induced by the
resonance of a SRR, whereas the optical band of a PC results from
multiple scattering, which thus arises at higher frequencies.  We are
therefore in presence of a periodic structure behaving somewhat as a
composite intermediate between a metamaterial and a photonic crystal.
One of the most topical subjects in photonics is the so-called all-angle-negative-
refraction (AANR), which was first described in \cite{zengerle87a}.
AANR allows one to focus light emitted by a point, onto an image, even through a
flat lens, provided that certain conditions for AANR are met, such
as convex isofrequency contours shrinking with frequency
about a point in the Brillouin zone \cite{luo02a}. In Fig. \ref{fig:aanrbis}, we show
such an effect for a perfectly conducting photonic crystal (PC)
in Fig. \ref{fig:aanrbis}(a).
 In order to achieve AANR, we choose a frequency on
the first dispersion curve (acoustic band) in Fig. \ref{fig:2D_Neumann}(c),
and we take its intersection with the light line $\Omega= \mid\kappa\mid$
along the $X\Gamma$ path. This means that we achieve negative group
velocity for waves propagating along the
$X\Gamma$ direction of the array, hence the rotation by an angle $\pi/4$
of every cell within the PC in panel (b) of Fig. \ref{fig:aanrbis}. This
is a standard trick in optics that has the effect of moving the origin of the
light-line dispersion to $X$ as, relative to the PC, the Bloch
wavenumber is along $X\Gamma$. This then creates optical effects due
to the interaction of the light-line with the acoustic branch, this
would be absent if $\Gamma$ were the light-line origin.

 The anisotropy of the effective material is reflected from
 coefficients $T_{11}=-5.53$ and $T_{22}=0.2946$. The same frequency
 of the first band is reachable at point $N$ of the Brillouin zone. By
 symmetry of the crystal, we would have $T_{11}=0.2946$ and
 $T_{22}=-5.53$. The resultant propagating waves would come from the
 superposition of the two effective media described
 above. Fig. \ref{fig:aanrbis}(b) illustrates this anisotropy as the
 source wave only propagates at the prescribed directions.

\begin{figure}
  \begin{center}
    \includegraphics[scale=0.8]{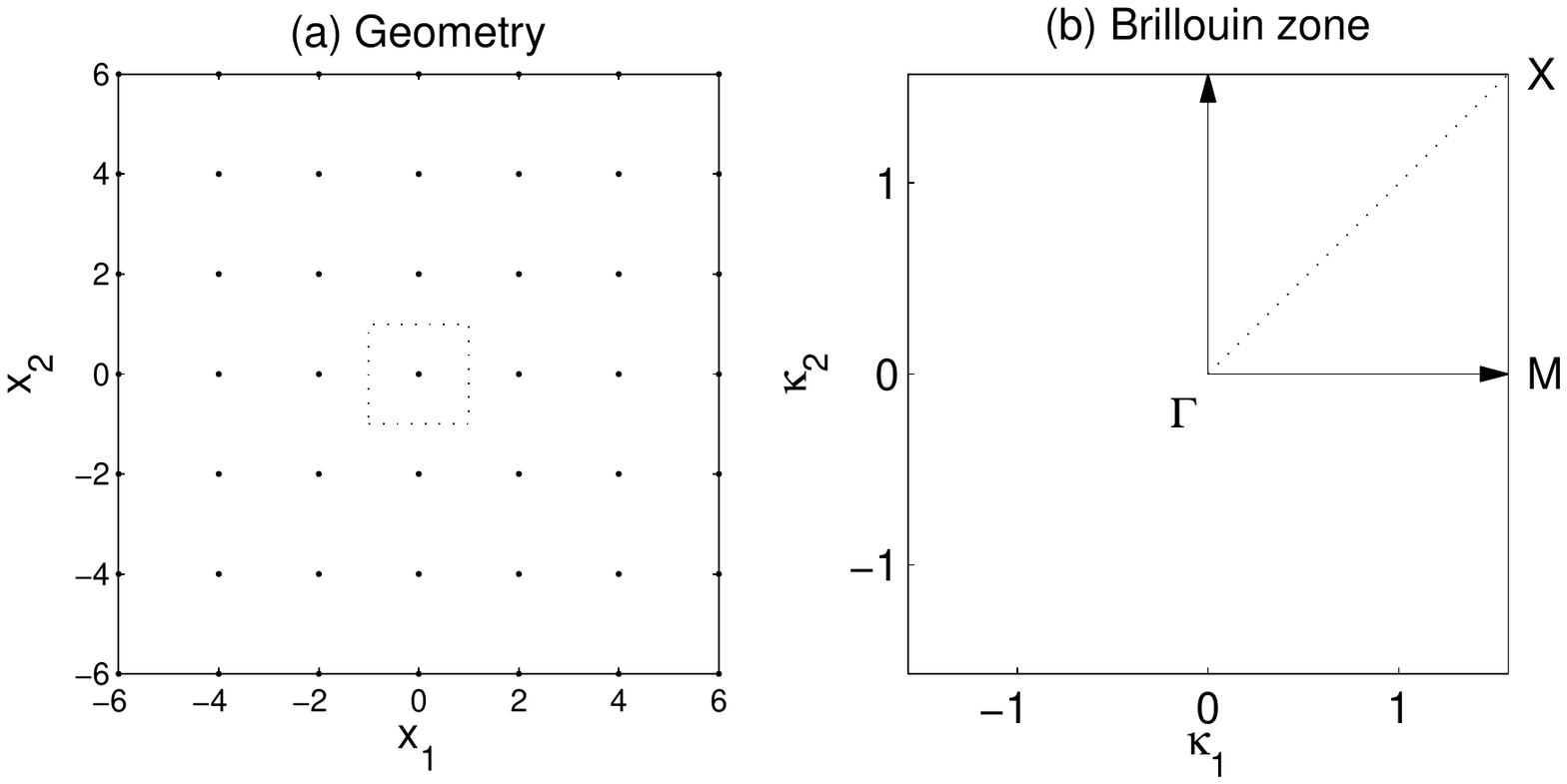}
   \end{center}
\vspace{-8cm}
\caption{For the two dimensional example we show the geometry of the
  doubly periodic simply supported plate (the dots represent the
  simple supports)  in panel (a) with the
  elementary cell shown by the dotted lines and in (b) the irreducible Brillouin
  zone with the lettering for wavenumber positions shown. 
Figure reproduced from Proceedings of the Royal Society \cite{antonakakis12a}.
}
\label{fig:2dgeom}
\end{figure}

\subsection{Kirchoff Love Plates}
\label{sec:Plate}
HFH is by no means limited to the Helmholtz operator. HFH is here
applied to flexural waves in two dimensions \cite{antonakakis12a} for
which the governing equation is a fourth order equation 
\beq
\nabla^4 u -\Omega^2 u = 0;
\eeq
 assuming constant material parameters. Such a thin plate can be
 subject to point, or line, constraints and these are common place in
 structural engineering. 

In two dimensions, only a few examples of constrained plates are available in the
literature: a grillage of line constraints as in \cite{mace81a} that is
effectively two coupled one dimensional problems, a periodic line array
of point supports \cite{evans07a} raises 
the possibility of Rayleigh-Bloch modes and for doubly periodic point
supports there are exact solutions by \cite{mace96a} (simply
supported points) and by \cite{movchan07c} (clamped points); the simply
supported case is accessible via Fourier series and we choose this as
an illustrative example that is of interest in its own right; it is
shown in figure \ref{fig:2dgeom}(a). In
particular the simply supported plate has a zero-frequency stop-band
and a non-trivial dispersion diagram. It is worth noting that
classical homogenization is of no use in this setting with a zero
frequency stop band. Naturally waves passing through periodically
constrained plates have many similarities with those of photonics in optics.

We consider a
double periodic array of points at $x_1=2n_1$, $x_2=2n_2$ where $u=0$
(with the first and second derivatives continuous) and so
the elementary cell is one in $\vert x_1\vert<1, \vert x_2\vert<1$ with
$u=0$ at the origin (see Figure \ref{fig:2dgeom});  Floquet-Bloch conditions are applied at the
edges of the cell.

\begin{figure}
  \begin{center}
    \includegraphics[scale=0.7]{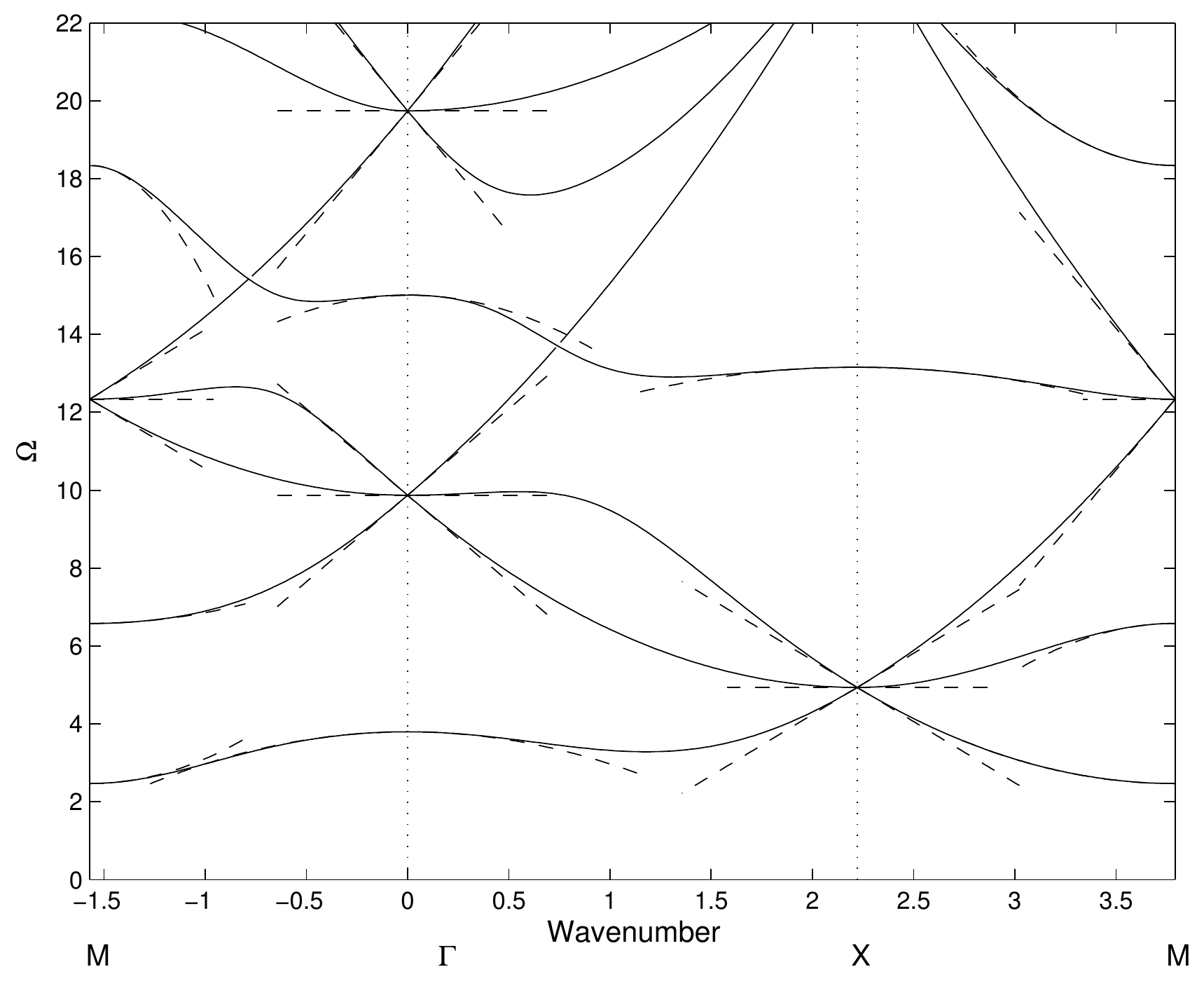}
   \end{center}
\vspace{-0.4cm}
\caption{The dispersion diagram for a doubly periodic array of point
  simple supports shown for the irreducible Brillouin zone of
  Fig. \ref{fig:2dgeom}. The figure shows the dispersion curves as solid lines. As dashed lines, the
  asymptotic solutions from the high frequency homogenization theory
  are shown. Figure reproduced from Proceedings of the Royal Society \cite{antonakakis12a}}
\label{fig:2Dplate}
\end{figure} 

Applying Bloch's theorem and Fourier series the displacement is
readily found  \cite{mace96a} as 
\beq
  u({\bf x})= \exp(i{\bm \kappa}\cdot{\bf x})\sum_{n_1,n_2}
  \frac{\exp(-i\pi{\bf N}\cdot{\bf x})}{[(\kappa_1-\pi
    n_1)^2+(\kappa_2-\pi n_2)^2]^2-\Omega^2},
\label{eq:2du}
\eeq
where ${\bf N}=(n_1,n_2)$, and enforcing the condition at the origin gives the dispersion relation
\beq
   D(\kappa_1,\kappa_2,\Omega)=\sum_{n_1,n_2}\frac{1}{[(\pi
     n_1-\kappa_1)^2 +(\pi n_2 -\kappa_2)^2]^2-\Omega^2}=0,
\label{eq:2d_dispersion}
\eeq 

 In this two dimensional example a
  Bloch wavenumber vector ${\bm\kappa}=(\kappa_1,\kappa_2)$ is used
  and the dispersion relation can be characterised completely by
  considering the irreducible Brillouin zone $\Gamma XM$ shown in figure
  \ref{fig:2dgeom}. 

 The dispersion diagram is shown in figure \ref{fig:2Dplate}; The singularities of the summand in equation (\ref{eq:2d_dispersion}) correspond to solutions within the cell
 satisfying the Bloch conditions at the edges, in some cases these
 singular solutions also satisfy the conditions at the support and are
 therefore true solutions to the problem, a similar situation occurs
 in the clamped case considered using multipoles in
 \cite{movchan07c}. Solid lines in figure \ref{fig:2Dplate} label curves
 that are branches of the dispersion relation, notable features are
 the zero-frequency stop-band and also crossings of branches at the
 edges of the Brillouin zone. Branches of the dispersion relation that
 touch the edges of the Brillouin zone singly fall into two
 categories, those with multiple modes emerging at a same standing wave frequency (such as
 the lowest branch touching the left handside of the figure at M) and
 those that are completely alone (such as the second lowest branch on
 the left at M).

The HFH theory can again be employed to find an effective PDE entirely
upon the long-scale that describes the behaviour local to the standing
wave frequencies and the details are in \cite{antonakakis12a}, the
asymptotics from the effective PDE are shown in Fig. \ref{fig:2Dplate}
as the dashed lines. 

\section{High-contrast homogenization}\label{kc}

Periodic media offer a convenient tool in achieving control of electromagnetic waves, due to 
their relative simplicity from the point of view of the manufacturing process, and due to the possibility of using 
the Floquet-Bloch decomposition for the analysis of the spectrum of the wave equation in such media. The latter issue has received a considerable amount of interest in the mathematical community, in particular from the perspective of the inverse problem: how to achieve a given spectrum and/or density of states for the 
wave operator with periodic coefficients by designing an appropriate periodic structure? While the Floquet-Bloch decomposition provides a transparent procedure for answering the direct question, it does not yield a straightforward way of addressing the inverse question posed above. 
  
  One possibility for circumventing the difficulties associated with the inverse problem is by viewing the given periodic structure as a high-contrast one, if this is possible under the values of the material parameters used. The idea of considering high-contrast composites within the context of homogenization appeared first in the work by Allaire \cite{Allaire}, which discussed the application of the two-scale convergence technique (Nguetseng \cite{Nguetseng}) to classical homogenization. A more detailed analysis of high-contrast composites, 
along with the derivation of an explicit formula for the related spectrum, was carried out in a major study by Zhikov \cite{Zhikov2000}. One of the obvious advantages in using high-contrast composites, or viewing a given composite as a high-contrast one, is in the mere existence of such formula for the spectrum. In the present section we focus on the results of the analysis of Zhikov, and on some more recent results for one-dimensional, layered, high-contrast periodic structures.

In order to get an as short as possible approach to the high-contrast theory, we consider the equation of electromagnetic wave propagation in the transverse electric (TE) polarisation, when the magnetic field has the form 
$(0,0,H),$ in the presence of sources with spatial density $f({\bf x}):$ 
\beq
-{\rm div}(\varepsilon^\eta)^{-1}\left({\bf x}/\eta\right)\nabla H({\bf x})
=\omega^2 H({\bf x})+f({\bf x}), \ \ \ \ {\bf x}\in\Omega\subset{\mathbb R}^2,
\label{TMeq}
\eeq
where we normalise the speed of light $c$ to 1 for simplicity, which amounts to taking
$\varepsilon_0\mu_0=1$ in section \ref{clashom}, and
where the magnetic permeability is assumed to be equal to unity throughout the medium
({\it i.e.} $\mu=\mu_0$), and the function 
$f({\bf x})$ is assumed to vanish outside some set that has positive distance to the boundary of $\Omega.$ 
The inverse dielectric permittivity tensor $(\varepsilon^\eta)^{-1}({\bf y})$ is assumed in this section, for simplicity, 
to be a scalar, taking values $\eta^\gamma I$ and $I,$ respectively, on  $[0,1]^2$-periodic open sets $F_0$ 
and $F_1,$ such that $\overline{F_0}\cup\overline{F_1}=
{\mathbb R}^2.$ Here $\gamma$ is a 
positive exponent representing a ``contrast'' between material properties of the two components of the 
structure that occupy the regions $F_0$ and $F_1.$ In what follows we also  assume that $F_0\cap[0,1]^2$ has a finite distance to the boundary of 
the unit cell $[0,1]^2,$ so that the ``soft'' component $F_0$ consists of disjoint ``inclusions'', spaced 
$[0,1]^2$-periodically from each other, while the ``stiff'' component $F_1$ is a connected subset of ${\mathbb R}^2.$
The matrix $\varepsilon^\eta$ represents the dielectric permittivity of the medium at a given point, however the analysis and conclusions of this section are equally applicable to acoustic wave propagation, which is the 
context we borrow the terms ``soft'' and ``stiff'' from. The assumed relation between the values of dielectric 
permittivity  $\varepsilon^\eta$  (in acoustics, between the ``stiffnesses'' ) on the two components of the 
structure is close to the setting of what has been described as ``arrow fibres'' in the physics literature on electromagnetics, see {\it e.g}
\cite{pcfbook}.

A simple dimensional analysis shows that if $\omega\sim 1$ then the soft inclusions are in resonance with the 
overall field if and only if $\gamma=2,$ which is the case we focus on henceforth. 

The above equation (\ref{TMeq}) describes the wave profile for a TE-wave in the cylindrical domain 
$\Omega\times{\mathbb R}$ domain, and it is therefore supplied with the Neumann condition $\partial H/\partial n=0$
\footnote{Neumann boundary conditions {\it i.e.} infinite conducting walls is a good model for metals in
microwaves, but much less so in the visible range of frequencies
wherein absorption by metals need be taken into account. Note also that in the TM polarization case,
when the electric field takes the form $(0,0,E)$, our analysis applies {\it mutatis mutandis} by
interchanging the roles of $\varepsilon$ and $\mu$, $H$ and $E$, and Neumann boundary
conditions by Dirichlet ones.}
on the 
boundary of the domain and with the Sommerfeld radiation condition  $\partial H/\partial\vert x\vert-{\rm i}\omega H=o(\vert x\vert^{-1})$ as $\vert x\vert\to\infty.$ 

In line with the previous sections, we apply the method of two-scale asymptotic expansions to the above problem, seeking the solution $H=H(x_1,x_2)=H({\bf x})$ in the form (see also (\ref{sebeq01} in Section \ref{clashom1})
\beq
H({\bf x})=H_0({\bf x},{\bf x}/\eta)+\eta H_1({\bf x},{\bf x}/\eta)+\eta^2 H_2({\bf x},{\bf x}/\eta)+...,
\label{twoscaleexp}
\eeq
where the functions involved are $[0,1]^2$-periodic with respect to the ``fast'' variable $y=x/\eta.$
Substituting the expansion (\ref{twoscaleexp}) into the equation (\ref{TMeq}) and rearranging the terms in the 
resulting expression in such a way that terms with equal powers of $\eta$ are grouped together,
we obtain a sequence of recurrence relations for the functions $H_k,$ $k=0,1,...,$ from which they are
obtained sequentially. The first three of these equations can be transformed to the following system of equations for the leading-order term $H^{(0)}({\bf x},{\bf y})=u({\bf x})+v({\bf x},{\bf y}),$ ${\bf x}\in\Omega,$ ${\bf y}\in[0,1]^2:$  
\beq
-{\rm div}\varepsilon_{\rm hom}^{-1}\nabla u({\bf x})=\omega^2\biggl(u({\bf x})
+\int_{F_0\cap[0,1]^2}v({\bf x},{\bf y})d{\bf y}\biggr)+f({\bf x}), \ \ \ {\bf x}\in\Omega,
\label{limit1}
\eeq
\beq
-\Delta_{\bf y} v({\bf x},{\bf y})
=\omega^2\bigl(u({\bf x})+v({\bf x},{\bf y})\bigr)
+f({\bf x}),\ \ \ \ y\in F_0\cap[0,1]^2,\ \ \ \ v({\bf x},{\bf y})=0,\ \ \  y\in{F_1}\cap[0,1]^2.
\label{limit2}
\eeq
These equations are supplemented by the boundary conditions for the function $u,$ of the same kind as in the 
problems with finite $\eta.$ For the sake of simplifying the analysis, we assume that those inclusions that 
overlap with the boundary of $\Omega$ are substituted by the ``main'', ``stiff'' material, where 
$(\varepsilon^\eta)^{-1}=I.$

In the equation (\ref{limit1}), the matrix $\varepsilon_{\rm hom}$ is the classical homogenization matrix for 
the perforated medium $\varepsilon F_1,$ see Section \label{clashom} above. However, the properties of the system 
(\ref{limit1})--(\ref{limit2}) are rather different to those for the perforated-medium homogenised limit, described 
by the equation 
$-{\rm div}\varepsilon_{\rm hom}^{-1}\nabla u({\bf x})=\omega^2u({\bf x})+f({\bf x}).$ As we shall see next, the two-scale structure of 
(\ref{limit1})--(\ref{limit2}) means that the description of the spectra of the problems (\ref{TMeq}) in the limit as 
$\eta\to0$ diverges dramatically from the usual moderate-contrast scenario.

The true value of the above limiting procedure is revealed by the statement of the convergence, as $\eta\to0,$ of the spectra of the original problems to the spectrum of the limit problem described above, see \cite{Zhikov2000} and 
by observing that the spectrum of the system (\ref{limit1})--(\ref{limit1}) is evaluated easily as follows. We write an eigenfunction expansion for $v({\bf x},{\bf y})$ as a function of $y\in{F_0}\cap[0,1]^2:$ 
\beq
v({\bf x},{\bf y})=\sum_{k=0}^\infty c_k({\bf x})\psi_k({\bf y}),
\label{vform}
\eeq
where $\psi_k$ are the (real-valued) eigenfunctions of the Dirichlet problem $-\Delta\psi_k=\lambda_k\psi_k,$  
$y\in{F_1}\cap[0,1]^2,$ arranged in the order of increasing eigenvalues $\lambda_k,$ $k=0,1,...$ and orthonormalised according to the conditions $\int_{{F_0}\cap[0,1]^2}\vert\psi_k({\bf y})\vert^2d{\bf y}=1,$ $k=0,1,...,$ and 
$\int_{{F_0}\cap[0,1]^2}\psi_k({\bf y})\psi_l({\bf y})d{\bf y}=0,$ $k\neq l,$ $k,l=0,1,...$ Substituting (\ref{vform}) into (\ref{limit2}), 
we find the values for the coefficients  $c_k,$ which yield an explicit expression for $v({\bf x},{\bf y})$ in terms of the function $u({\bf x}):$ 
\[
v({\bf x},{\bf y})=\bigl(\omega^2u({\bf x})+f({\bf x})\bigr)
\sum_{k=0}^\infty\Bigl(\int_{{F_0}\cap[0,1]^2}\psi_k({\bf y})d{\bf y}\Bigr)(\lambda_k-\omega^2)^{-1}\psi_k({\bf y}).
 \] 
Finally, using the last expression in the first equation in (\ref{limit1}) yields an equation for the function $u$ only:
\beq
-{\rm div}\varepsilon_{\rm hom}^{-1}\nabla u({\bf x})=\beta(\omega^2)\bigl(u({\bf x})+\omega^{-2}f({\bf x})), \ \ \ {\bf x}\in\Omega,
\label{homog1}
\eeq
where 
the function $\beta,$ which first appeared in the work \cite{Zhikov2000}, is given by
\beq
\beta(\omega^2)=\omega^2\biggl(1+\omega^2
\sum_{k=0}^\infty\Bigl(\int_{{F_0}\cap[0,1]^2}\psi_k({\bf y})d{\bf y}\Bigr)^2(\lambda_k-\omega^2)^{-1}\biggr).
\label{betaformula}
\eeq

\begin{center}
\begin{figure}[h]
\centering
\includegraphics[scale=0.5]{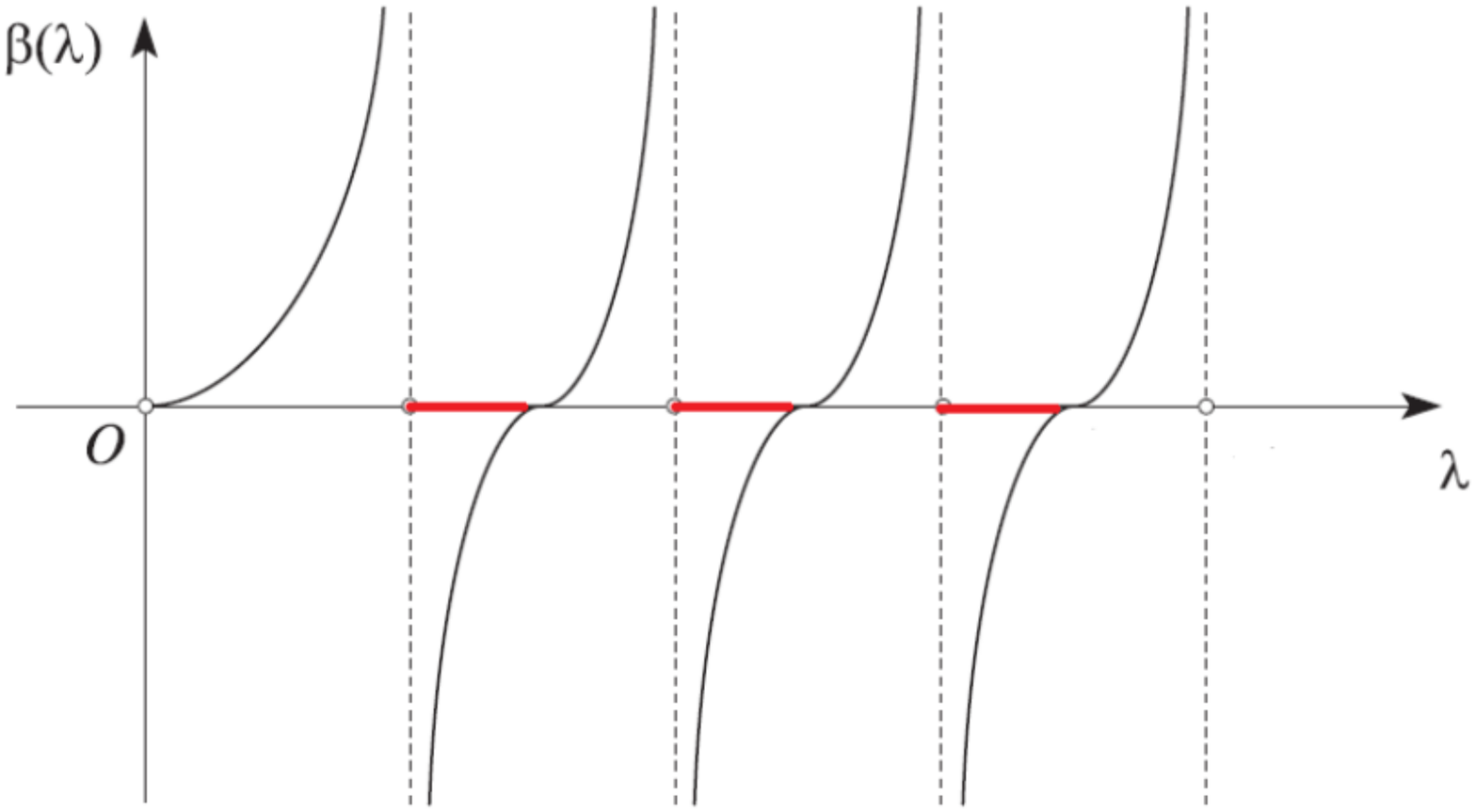}
\caption{The plot of the function $\beta$ describing the spectrum of the problem (\ref{limit1})--(\ref{limit2}) subject to the boundary conditions. 
The stop bands for the problem in the whole space ${\mathbb R}^2$ are indicated by the red intervals of the horizontal axis. The spectra of the problems (\ref{TMeq}) considered in the whole space converge, as $\eta\to0,$ to the closure of the complement of the union of the red intervals in the positive semiaxis.}
\label{shane_figure} 
\end{figure}
\end{center}   

The equation (\ref{homog1}) is supplemented by appropriate boundary conditions and/or conditions at infinity, 
which are inherited from the $\eta$-dependent family, {\it i.e.} the Neumann condition at the boundary points 
${\bf x}\in\partial\Omega$ and the radiation condition when $\vert {\bf x}\vert\to\infty.$  Clearly, the spectrum of this limit 
problem consists of those values of $\omega^2$ for which $\beta(\omega^2)$ is in the spectrum of the 
operator generated by the differential expression $-{\rm div}\varepsilon_{\rm hom}^{-1}\nabla$ subject to the same boundary conditions. For example, for the problem in the whole space ${\mathbb R}^2$ 
(describing the behaviour of TE-waves in a 3D periodic structure that is invariant in one specified direction) 
this  procedure results in a band-gap spectrum shown in Fig. \ref{shane_figure}. The end points of each pass band are found by 
a simple analysis of the formula (\ref{betaformula}): the right ends of each pass band are given by those 
eigenvalues $\lambda_k$ of the Dirichlet Laplacian on the inclusion ${F_0}\cap[0,1]^2$ that possess at least one eigenfunction with non-zero integral over ${F_0}\cap[0,1]^2$ (otherwise the corresponding term in 
(\ref{betaformula}) vanishes), while the left ends of the pass bands  are given by solutions to the polynomial equation of infinite order $\beta(\omega^2)=0.$ These points have a physical interpretation as eigenvalues of the 
so-called electrostatic problem on the inclusion, see \cite{Zhikov2005}. 

As in the case of classical, moderate-contrast, periodic media, the fact of spectral convergence offers significant computational advantages over tackling the equations (\ref{TMeq}) directly: as $\eta\to0$ the latter becomes increasingly demanding, while the former requires a single numerical procedure that serves all $\eta$ once the homogenised matrix $\varepsilon_{\hom}$ and several eigenvalues $\lambda_k$ are calculated.  A significant new feature, however, as compared to the classical case, 
is the fact of an infinite set of stop bands opening in the limit as $\eta\to,$ which are easily controlled by the explicit description of the band endpoints. This immediately yields a host of applications of the above results for the design 
of  band-gap devices with prescribed behaviour in the frequency interval of interest.  

The theorem on spectral convergence for problems described by the equation (\ref{TMeq}) is proved in \cite{Zhikov2000} under the assumption of connectedness 
of the domain $F_1$ occupied by the ``stiff'' component, via a variant of the extension procedure from $F_1$ to the whole of ${\mathbb R}^2$ for function sequences whose energy scales as $\eta^{-2}$ (or, equivalently, finite-energy 
sequences for the operator prior to the rescaling ${\bf x}/\eta={\bf y}$). In the more recent works \cite{CCG}, \cite{CC}, this assumption is dropped in a theorem about spectral convergence for a general class of high-contrast  operators, via a version of the two-scale asymptotic analysis akin to 
(\ref{twoscaleexp}), for the Floquet-Bloch components of the resolvent of the original family of operators following the re-scaling
${\bf x}/\eta={\bf y}.$  
In particular, in \cite{CCG} a one-dimensional high-contrast model is analysed, which in 3D corresponds to a 
stack of dielectric layers aligned perpendicular to the direction of the magnetic field. Here the procedure described above for the 2D grating fails to yield a satisfactory 
limit description as $\eta\to0,$ {\it i.e.} a description where the spectra of problems for finite $\eta$ converge to the spectrum of the limit problem described by the system (\ref{limit1})--(\ref{limit2}) as $\eta\to0.$ A more refined analysis  of the structure of the related $\eta$-dependent family results in a statement of convergence to the set described by 
the inequalities
\beq
-1\le\frac{1}{2}(\alpha-\beta+1)\sqrt{\lambda}\sin\Bigl(\sqrt{\lambda}(\alpha-\beta)\Bigr)
+\cos\Bigl(\sqrt{\lambda}(\alpha-\beta)\Bigr)\le1.
\label{limitspectrumformula}
\eeq
where $\alpha$ and $\beta$ denote the end-points of the inclusion in the unit cell, {\it i.e.} 
$F_0\cap[0,1]^2=(\alpha,\beta)\times[0,1].$

\begin{center}
\begin{figure}[h]
\centering
\includegraphics[scale=0.5]{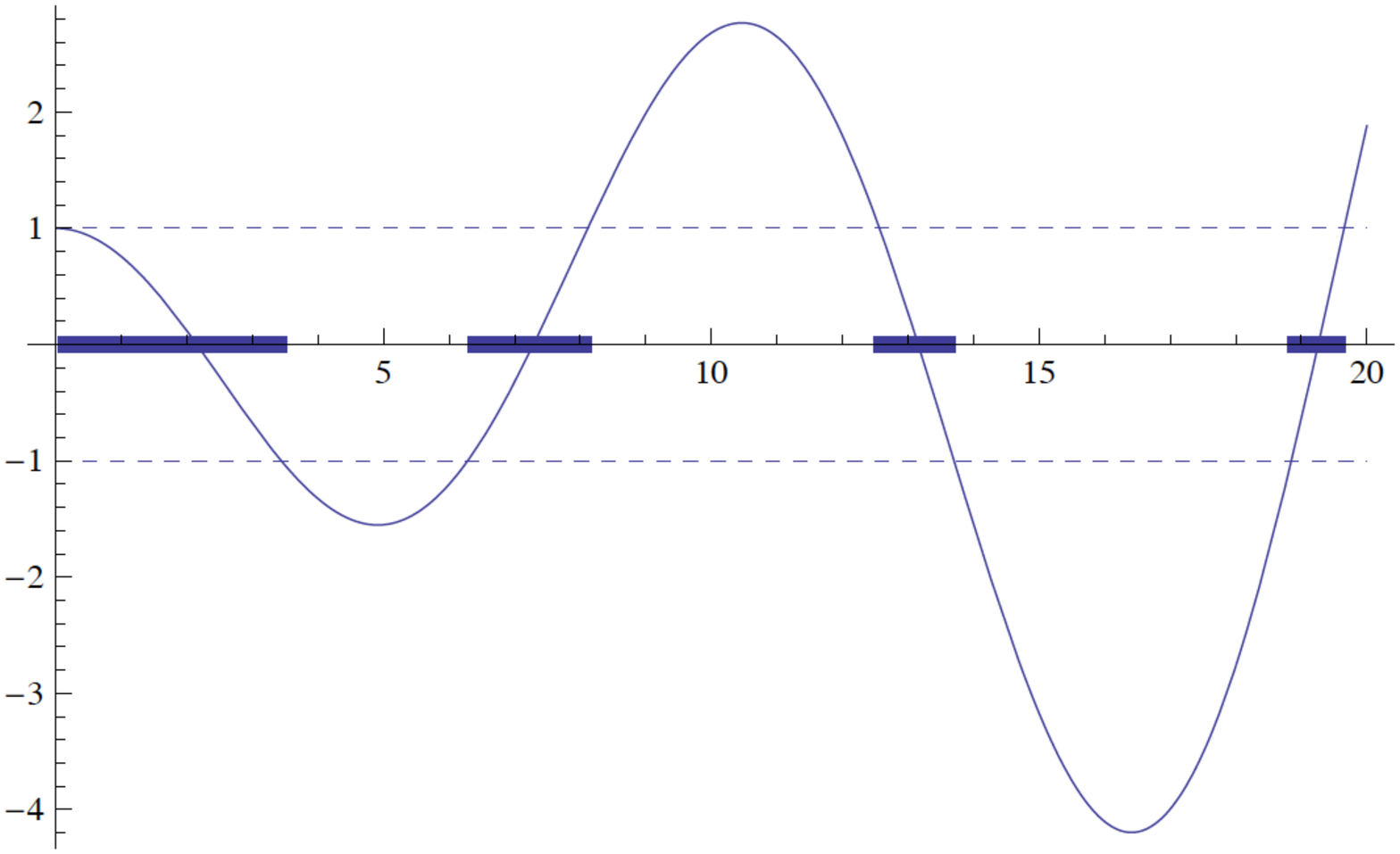}
\caption{The square root of the limit spectrum for a 1D high-contrast periodic stack, in TE polarisation. 
The oscillating solid line is the graph of the function
$f(\omega)=\cos(\omega/2)-\omega\sin(\omega/2)/4$ in (\ref{limitspectrumformula}) with $\alpha=1/4,$ 
$\beta=3/4.$ The square root of the spectrum is the union of the intervals indicated by bold lines. }
\label{limitspectrum} 
\end{figure}
\end{center}

Similarly to  the spectrum of the 2D high-contrast problem, described by the function $\beta,$ the limit spectrum of the 1D problem has a band-gap structure, shown in Fig. \ref{limitspectrum}, however the description of the location of the bands is different  in 
that it is no longer obtained from the inequality $\beta>0,$ where $\beta$ is the 1D analogue of (\ref{betaformula}).
Importantly, the asymptotic behaviour of the density of states function as $\eta\to0$ is also very different 
in the two cases. One can show that the family of resolvents for the problems (\ref{TMeq}) converges, up to a 
suitable unitary transformation, to the resolvent of a certain operator whose spectrum is given exactly by 
(\ref{limitspectrumformula}), see \cite{CC}. The rate of convergence is rigorously shown to be $O(\eta),$ as is anticipated by the expansion (\ref{twoscaleexp}).

The above 1D result is generalised to the case of an oblique incidence of an electromagnetic wave on the same 3D layered structure. Suppose that $x_2$ is the coordinate across the stack. Then, assuming for simplicity that the wave vector $(\varkappa,0,0)$ 
is parallel to the direction $x_1,$ it can be shown that all three components of the magnetic field are non-vanishing, with the magnetic component $H=H_3$ satisfying the equation 
\[
-\Bigl((\varepsilon^\eta)^{-1}(x/\eta)H'(x)\Bigr)'=\Bigl(\omega^2-(\varepsilon^\eta)^{-1}(x/\eta)\varkappa^2\Bigr)H(x),
\]
subject to the same boundary conditions as before.
The modified limit spectrum for this family is given by those $\omega^2$ for which 
({\it cf.} (\ref{limitspectrumformula}))
\beq
-1\le\frac{1}{2}(\alpha-\beta+1)\biggl(\omega-\frac{\varkappa^2}{\omega}\biggr)\sin\Bigl(\sqrt{\lambda}(\alpha-\beta)\Bigr)
+\cos\Bigl(\sqrt{\lambda}(\alpha-\beta)\Bigr)\le1,\ \ \ \ \omega>0,
\label{limitspectrumformulamoodified}
\eeq
where, as before, $\alpha$ and $\beta$ describe the ``soft" inclusion layer in the unit cell, see \cite{CCG}.
The set of $\omega$ described by the inequalities (\ref{limitspectrumformulamoodified})
is similar to that shown in Figure \ref{limitspectrum}, the only significant difference between
the two cases being a low-frequency gap opening near $\omega=0$ for
(\ref{limitspectrumformulamoodified}).

\section{Conclusion and further applications of grating theory}
To conclude this chapter, we would like to stress that advances in homogenization theory over the past
forty years have been fuelled by research in composites \cite{milton02a}.
The philosophy of the necessity for rigour expressed by Lord Rayleigh in 1892 concerning the Lorentz-Lorenz
equations (also known as Maxwell-Garnett formulae) can be viewed as the foundation act of homogenization:
`In the application of our results to the electric theory of light we contemplate a medium interrupted by spherical, or
cylindrical, obstacles, whose inductive capacity is different from that of the undisturbed medium. On the other hand,
the magnetic constant is supposed to retain its value unbroken. This being so, the kinetic energy of the electric
currents for the same total flux is the same as if there were no obstacles, at least if we regard the wavelength as infinitely great.'
In this paper, John William Strutt, the third Lord Rayleigh \cite{rayleigh}, was able to solve Laplace's equation
in two dimensions for rectangular arrays of cylinders, and in three-dimensions for cubic lattices of spheres.
The original proof of Lord Rayleigh suffered from a conditionally convergent sum in order to compute the
dipolar field in the array. Many authors in the theoretical physics and applied mathematics communities
proposed extensions of Rayleigh's method to avoid this drawback. Another limit of Rayleigh's
algorithm is that it does not hold when
the volume fraction of inclusions increases. So-called multipole methods have
been developed in conjunction with lattice sums
in order to overcome such obstacles, see {\it e.g.} \cite{mmp}
for a comprehensive review of these methods.
In parallel to these developments, the quasi-static
limit for gratings has been the subject
of intensive research, one might
cite \cite{ross1} and \cite{petit}
for important contributions in
the 1980s, and \cite{popov}
for a comprehensive review
of the modern theory of
gratings, including a close
inspection of homogenization
limit.

\begin{figure}
\begin{center}
\scalebox{0.2}{\includegraphics{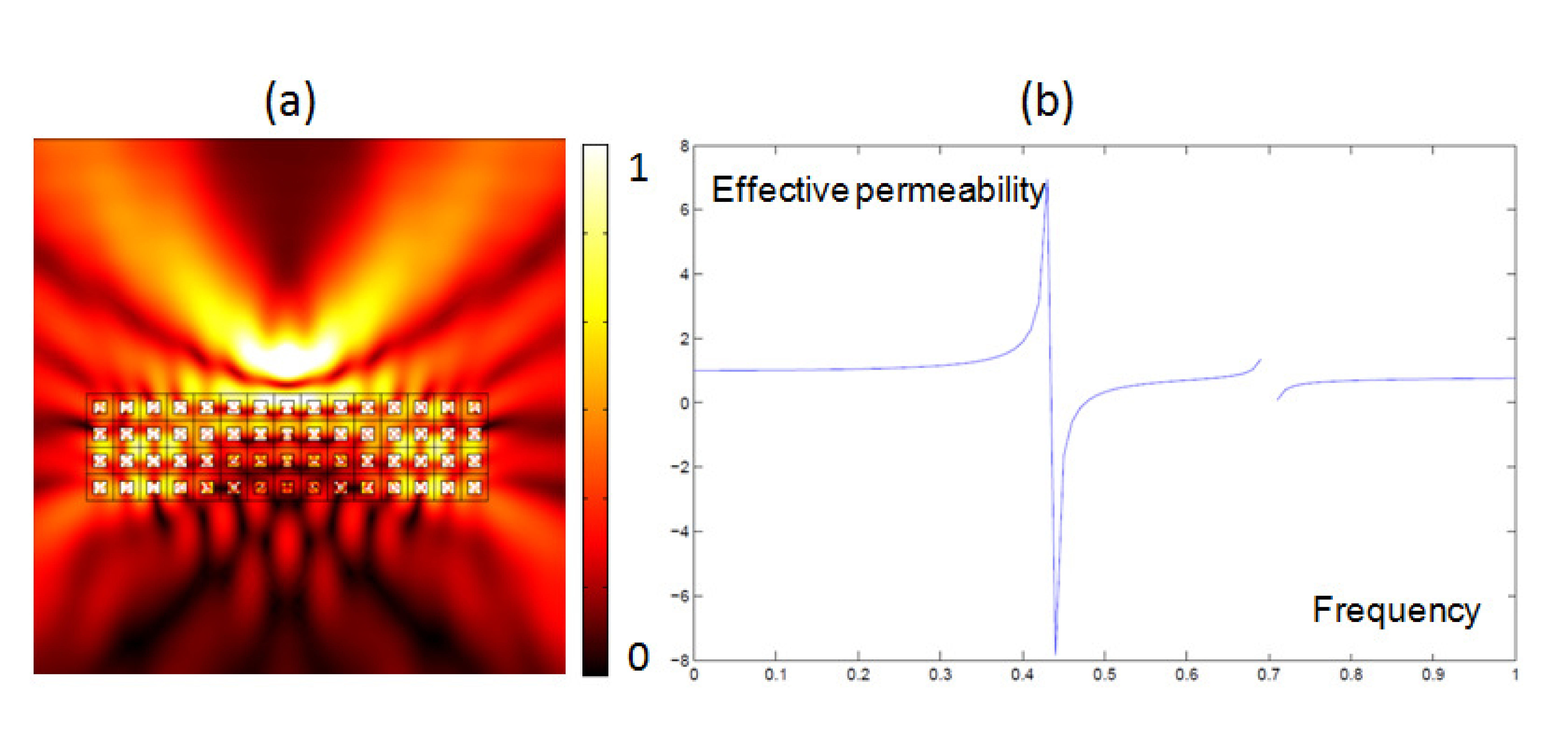}}
 \caption{Superlens application of grating:
(a) A time harmonic source at frequency $0.473$ displays an image through a square array of
square inclusions; (b) Effective magnetism versus frequency
using (\ref{effmag}) for square inclusions of relative permittivity $100$ with sidelength
$a=0.5d$ in matrix of relative permittivity $1$ (grating pitch $d=0.1$);
Negative values of the effective magnetism are in the frequency region $[0.432,0.534]$.
}
\label{figlast}
\end{center}
\end{figure}

Interestingly, in the pure mathematics community, Zhikov's work on high-contrast homogenization
\cite{Zhikov2000} has had important applications
in metamaterials, with the interpretation of his homogenized
equations in terms of effective magnetism first
put forward by O'Brien and Pendry \cite{obrienpendry},
and then by Bouchitt\'e and Felbacq
\cite{bouchitte}, although these
authors did not seem to be
aware at that time of
Zhikov's seminal
paper \cite{Zhikov2000}.
In order to grasp the
physical importance
of (\ref{homog1})-(\ref{betaformula}),
we consider the case of square inclusions of sidelength $a=d/2$, where
$d$ is the pitch of a bi-periodic grating. The eigenfunctions are
$\psi_{nm}({\bf y})=2\sin(n\pi y_1)\sin(n\pi y_2)$
in (\ref{betaformula})
and the corresponding eigenvalues
are $k^2_{nm}=\pi^2(n^2+m^2)$.
The right-hand side in the
homogenized equation (\ref{homog1})
can then be interpreted
in terms of effective magnetism:
\begin{equation}
\mu_{hom}(k)=1+\frac{64a^2}{\pi^4}\sum_{(n,m)odd} \frac{k^2}{n^2m^2(k^2_{nm}/a^2-k^2)} \; .
\label{effmag}
\end{equation}
This function can be computed numerically for instance with Matlab
and demonstrates that negative values can be achieved for
$\mu_{hom}$ near resonances, see Fig. \ref{figlast}(b).
This allows for superlensing via negative refraction,
as shown in Fig. \ref{figlast}(a).

Finally, we would like to point out that high-order homogenization
techniques \cite{kirill2004} suggest that most gratings display
some artificial magnetism and chirality when the wavelength
is no longer much larger than the periodicity \cite{boriseb}.
We hope we have convinced the reader that there is a
whole new range of physical effects in gratings which
classical, high-frequency and high-contrast
homogenization theories can capture. 

\let\cleardoublepage\clearpage

\renewcommand\bibname{\normalsize{\hspace{12 pt}References:}}

\end{document}